\documentclass[aip,jcp,amsmath,amssymb,reprint,longbibliography,twocolumn,floatfix]{revtex4-1}
\usepackage{graphicx} 
\usepackage{epstopdf}
\usepackage{xcolor}
\usepackage[colorlinks=False,citecolor=blue]{hyperref}%
\usepackage{multirow}
\usepackage{float}
\usepackage{enumitem}

\usepackage[normalem]{ulem}

\begin{document}


\title{Polymer brush-induced depletion interactions and clustering of membrane proteins}

\author{Anvy Moly Tom}
 \affiliation{Korea Institute for Advanced Study, Seoul 02455, Korea}
\author{Won Kyu Kim}%
\email{wonkyukim@kias.re.kr}
\affiliation{Korea Institute for Advanced Study, Seoul 02455, Korea}

\author{Changbong Hyeon}
\affiliation{Korea Institute for Advanced Study, Seoul 02455, Korea}

\date{\today}
\begin{abstract}
	We investigate the effect of mobile polymer brushes on proteins embedded in biological membranes by employing both Asakura-Oosawa type of theoretical model and coarse-grained molecular dynamics simulations.  
	The brush polymer-induced depletion attraction between proteins changes non-monotonically with the size of brush. 	
	The depletion interaction, which is determined by the ratio of protein size to the grafting distance between brush polymers, increases linearly with brush size as long as the polymer brush height is shorter than the protein size. When the brush height exceeds the protein size, however, the depletion attraction among proteins is slightly reduced.   
	We also explore the possibility of brush polymer-induced assembly of a large protein cluster, which can be related to one of many molecular mechanisms underlying recent experimental observations of integrin nanocluster formation and signaling. 
\end{abstract}

\maketitle

\section*{Introduction}
In 1950s, Asakura and Oosawa (AO) proposed a simple theoretical model to explain the interaction of entropic origin between colloidal particles immersed in a solution of macromolecules \cite{asakura1954JCP,Asakura58JPS}, which is of great relevance to our understanding of organization and dynamics in cellular environment.  
According to the AO theory, rigid spherical objects immersed in the solution of smaller hard spheres representing the macromolecules are expected to feel fictitious attraction, termed depletion force. 
While the interaction energy of the system remains unchanged, the spherical objects can be attracted to each other.   
Bringing the large spherical objects into contact  
can increase the free volume accessible to the smaller hard spheres comprising the medium, and hence increasing the total entropy of the hard sphere system ($\Delta S>0$). 
The free energy reduction due to the gain in entropy is 
\begin{align}
\Delta F_\text{HS}=-T\Delta S=-\left(\frac{3}{2}\lambda+1\right)\phi k_\text{B}T, 
\label{eqn:dF3}
\end{align}
where $\lambda$ is the size ratio of large to small hard spheres, and $\phi$ is the volume fraction of small spheres comprising the surrounding medium \cite{asakura1954JCP,Marrenduzzo06JCB,jeon2016Softmatter}.   
For a fixed value of $\phi$, 
the disparity in size between colloidal particles (large spheres) and macromolecular depletants (small spheres), characterized with the parameter $\lambda$, is the key determinant of the magnitude of depletion free energy \cite{Kang15PRL}. 
The effect of crowding environment on the aggregation of colloidal particles becomes substantial when $\lambda\gg 1$. 
The cellular environment is highly crowded, such that 30~\% of cytosolic medium is filled with macromolecules, rendering the interstitial spacing between macromolecules comparable to the average size of proteins $\sim$ 4 nm \cite{phillips2009physical}. 
More specifically, this volume fraction of \emph{E. coli} mixture is contributed by 11~\% of ribosome, 11 \% of RNA polymerase, and 8~\% of soluble proteins \cite{Roberts2011PLOSCOMPBIO}. 
In the cellular environment, the depletion force is one of the fundamental forces of great importance.

\begin{figure*}[ht]
	\includegraphics[width=0.75\linewidth]{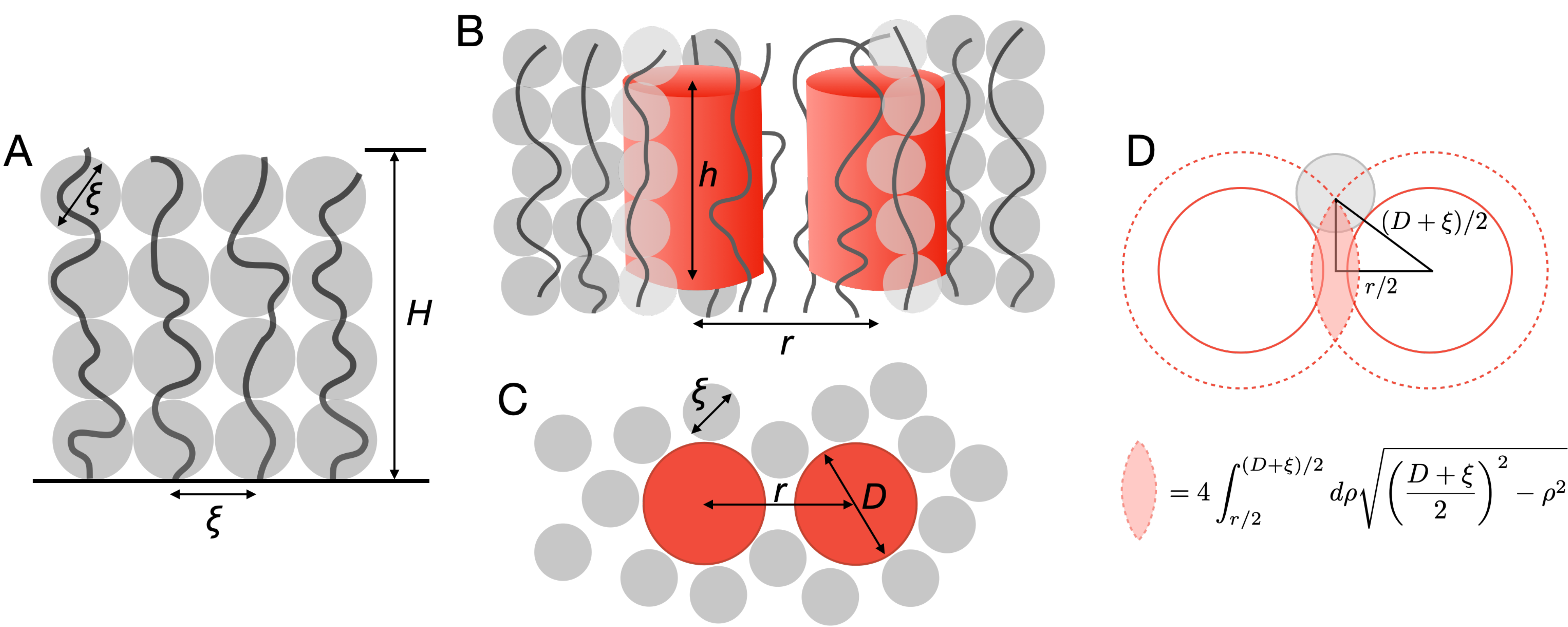}
	\caption{Brush-induced depletion interactions. 
	(A) Illustration of brush polymers, each of which is organized into a string of blobs of size $\xi$ above the surface.  
	(B) Two cylindrical inclusions (red) separated by distance $r$ surrounded by brush polymers (grey). 
	(C) Top view of (B). The lateral dimension of brush polymer $\xi$ corresponds to the size of a blob depicted with grey sphere (see (A)).  
	(D) Diagram to calculate the brush-induced depletion interaction between the two cylindrical objects. 
	The area inside the dashed line, corresponding to $2\pi[(D+\xi)/2]^2-A_\text{overlap}(r)$ in Eq.\ref{eqn:V}, is the area inaccessible to the blob of polymer brush of size $\xi$. 
	The shaded region in pale red is the overlapping area of the two discs of radius $(D+\xi)/2$, separated by the distance $r$. 
	\label{fig:AO_depletion}}
\end{figure*}

The basic principle of AO theory on rigid bodies with spherical symmetry is straightforward; however, application of the idea to the repertoire of biological and soft materials requires quantitative assessment of \emph{entropy}, which is nontrivial especially when crowders are characterized with non-spherical shape and/or with polydispersity \cite{Onsager49AnnNYAS,Dijkstra94PRL,mao1995PRL,mao1997JCP,Kang15JACS} and when the system is under a special boundary condition \cite{PhysRevLett.72.582,Dinsmore98PRL}.  
For the past decades, there has also been much interest toward understanding of the effects of crowding in biology \cite{Minton00COSB,mcguffee2010PLOS,ellis2001TIBS,sapir2015COCI,van1999EMBO,Marrenduzzo06JCB}, which includes crowding-induced structural transitions in disordered chiral homopolymers \cite{Snir2005Science,Kudlay09PRL}, protein/RNA folding \cite{Zhou08ARB,Elcock10COSB,Cheung05PNAS,Pincus08JACS,Kilburn10JACS,Denesyuk11JACS,PhysRevX.9.041035,Soranno2014PNAS}, gene regulation through DNA looping \cite{Li09NaturePhysics}, genome compaction \cite{Kim11PRL}, efficient search of proteins for targets on DNA \cite{Brackley13PRL}, and molecular motors \cite{Reddy2017Traffic,nettesheim2020NatPhys}. 
Further, it is worth mentioning a series of effort to understand the dynamics of active matter in the language of depletion forces~\cite{dzubiella2003depletion,angelani2011PRL,sanchez2012Nature,ni2015PRL,huber2018science}.

Besides the examples of depletion force-induced dynamics that all occur in three dimensional space, the AO theory can be extended to lateral depletion effects on the objects whose motion is confined in flat surfaces \cite{sintes1997BJ,suda2020lateral}. 
For biological membrane where the area fraction of membrane-embedded proteins is as high as 15 -- 30~\%, the formation of protein clusters or nano- or micro-domains \cite{Paszak14Nature,bakker2012PNAS,selhuber2008BJ,van2010PNAS,garcia2014JCS} is of great relevance to understanding the regulation of biological signal transduction and cell-to-cell communication.  
Although other physical mechanisms are still conceivable, lateral depletion interactions between membrane embedded proteins can arise from the fluctuations of lipids \cite{sintes1997BJ,soubias2015BJ,kusumi1982Biochem} or other polymer-like components comprising fluid membrane \cite{kim2006macromolecules,spencer2021macromolecules}, contributing to protein--protein attraction and clustering. 
In this context, the formation of integrin nanodomain which enables cell-to-cell communications via signaling \cite{williams1994TCB,kornberg1992JBC,paszek2009PLOS,cheng2020SciAdv},  
particularly, the bulky glycocalyx-enhanced integrin clusterings and the associated signaling-induced cancer metastasis observed by Paszek \emph{et al.} \cite{Paszak14Nature} make the brush polymer-induced depletion interaction between membrane proteins and their clustering a  topic of great relevance to investigate.  
 
In this paper, we study the lateral depletion interactions between rigid inclusions embedded in the mobile polymer brushes in 2D surface in the spirit of the AO theory in its simplest form. 
We compare the results from our simulations with our theoretical predictions. 
By analyzing the distribution of brush polymer-enhanced protein clusters obtained from our simulations, we attempt to link the brush-size dependent populations of giant protein clusters with the strength of signal transduction observed in Paszek \emph{et al.}'s measurement.

\section*{Theory: Brush-induced lateral depletion interactions}
As illustrated in Fig.~\ref{fig:AO_depletion}A, 
we consider flexible polymer brushes, each consisting of $N+1$ monomers of size (diameter) $b$. 
One end of individual chain is grafted to the surface but is free to move. 
If the grafting density $\sigma$ is large enough to satisfy $\sigma R_F^2>1$ \cite{deGennes1980Macromolecules,liu2017Macromolecules,liu2018JCP} or equivalently if the grafting distance ($\xi$) is smaller than $R_F=bN^{3/5}$, i.e., $\xi<R_F$, 
where $R_F$ is the Flory radius of the polymer in good solvent, 
each polymer reorganizes into a string of self-avoiding blobs due to excluded volume interactions with the neighboring polymers, forming a polymer brush of height $H$ where $N/g$ blobs of size $\xi$ consisting of $g$ segments fill the space above the surface (Fig.~\ref{fig:AO_depletion}A) \cite{deGennes1980Macromolecules}.
In this case, the grafting density $\sigma=N_\text{b}/A$, the number of polymer chains ($N_\text{b}$) grafted on an area $A$,  is related to the blob size (or the grafting distance) as $\sigma\simeq 1/\xi^2$. 
It is straightforward to show using the blob argument that the brush height $H$ scales with $N$ and $\sigma$ as \cite{Alexander77JP,deGennes1980Macromolecules,rubinstein2003polymer}
\begin{align}
H=N\sigma^{1/3}b^{5/3}. 
\label{eqn:brush_height}
\end{align}

\begin{figure}[t]
	\includegraphics[width=0.7\linewidth]{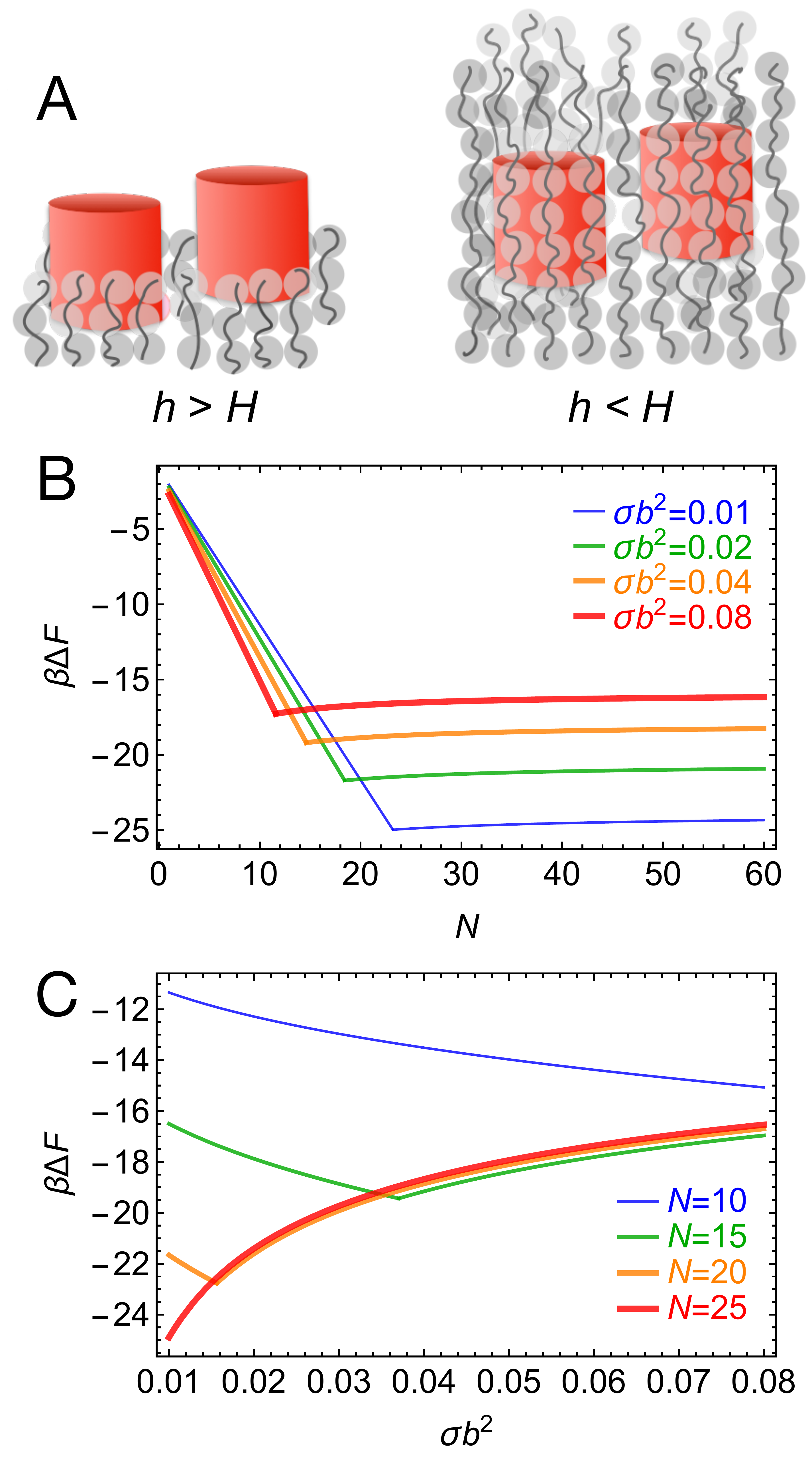}
	\caption{(A)~Two different cases of brush-induced depletion interaction:  
	$h>H$ (left), and $h<H$ (right). 
	(B), (C)~Free energy gain due to brush-induced depletion interaction. 
	Eq.~\eqref{eqn:dF2} was calculated as a function of $N$ for varying $\sigma$ (B), and as a function of grafting density ($\sigma$) for varying $N$ (C), with a cylindrical inclusion at fixed diameter $D=5b$ and height $h=5b$. 
	 \label{fig:height}}
\end{figure}

Our interest is in the lateral depletion force between two cylindrical inclusions  embedded in the polymer brush system, when the two inclusions, constrained to move in $xy$ plane, are separated by a fixed distance $r$ (Fig.~\ref{fig:AO_depletion}B, C). 
In the presence of the cylindrical inclusions, the volume accessible to the individual polymer chains is determined as follows, depending on $r$. 
\begin{widetext}
\begin{align}
V(r)=
\begin{cases}
AH-\left[2\pi\left(\frac{D+\xi}{2}\right)^2-A_\text{overlap}(r)\right]q(h,H), &\text{for }D\leq r\leq D+\xi \\
AH-2\pi\left(\frac{D+\xi}{2}\right)^2q(h,H), & \text{for }r > D+\xi. 
\end{cases}
\label{eqn:V}
\end{align}    
\end{widetext}
Here, $A_\text{overlap}(r)$ is the overlapping area between two circular discs of radius $(D+\xi)/2$, the region demarcated in pale red in Fig.~\ref{fig:AO_depletion}D, is 
\begin{align}
A_\text{overlap}(r)=4\int_{r/2}^{(D+\xi)/2}\left[\left(\frac{D+\xi}{2}\right)^2-\rho^2\right]^{1/2}\text{d}\rho.  
\end{align}
This is maximized when $r=D$, and its value can be written in terms of the area defined by the square of grafting distance, $\xi^2$, multiplied with a dimensionless factor $\chi(\lambda_\text{br})$,  
\begin{align}
A_\text{overlap}(D)&=\xi^2\underbrace{(1+\lambda_\text{br})\int^1_{\frac{\lambda_\text{br}}{1+\lambda_\text{br}}}(1-x^2)^{1/2}\text{d}x}_{\equiv\chi(\lambda_\text{br})}. 
\end{align}
where 
\begin{align}
\chi(\lambda_\text{br})
&=\frac{1}{2}
\left[(1+\lambda_\text{br})^2\cos^{-1}{\left(\frac{\lambda_\text{br}}{1+\lambda_\text{br}}\right)}-\lambda_\text{br}\sqrt{1+2\lambda_\text{br}}\right]\nonumber\\
&\simeq\begin{cases} 
\frac{\pi}{4}+\frac{\pi-2}{2}\lambda_\text{br}+\mathcal{O}(\lambda_\text{br}^2),
 &\mbox{for } \lambda_\text{br}\ll 1 \\ 
\frac{2\sqrt{2}}{3}\sqrt{\lambda_\text{br}}, & \mbox{for } \lambda_\text{br}\gg 1, \nonumber
\end{cases}
\end{align}
is a monotonically increasing function of $\lambda_\text{br}=D/\xi\simeq D\sqrt{\sigma}$, the ratio of the diameter of the inclusions to the grafting distance (or the blob size). 
Next, the function $q(h,H)\equiv H\Theta(h-H)+h\Theta(H-h)$, defined with the step function, signifies (i) $q(h,H)=H$ when the brush height ($H$) is shorter than the height of the inclusion ($h$) ($H<h$); and (ii) $q(h,H)=h$ when the brush is grown over the inclusion ($H>h$) (see Fig.\ref{fig:height}A).   
It is assumed that when $H>h$ the volume above the inclusions, $A\times(H-h)$, 
is fully accessible to the polymer chains, which is a reasonable assumption when $H\gg h$.  
Furthermore, under an assumption of no correlation between the polymer chains, 
the partition function for the brush system in the presence of the 2D inclusions separated by $r$ is 
$Z(r)=[V(r)]^{N_\text{b}\times (N+1)}$, where $N_\text{b}$ is the number of polymers consisting the brush.  
The thermodynamic equilibrium is attained by maximizing the total entropy of the system or minimizing the free energy $\beta F(r)=-\log{Z(r)}=-N_\text{b}(N+1)\log{V(r)}$. 
The gain in free energy due to depletion attraction can be obtained by taking the difference after and before the inclusions are in full contact with each other as $\beta\Delta F=\beta F(D)-\beta F(r\geq D+\xi)$  (see Appendix A for an alternative derivation using the depletion force): 
\begin{align}
-\beta&\Delta F=N_b(N+1)\log{\frac{V(D)}{V(r\geq D+\xi)}}\nonumber\\
&=N_b(N+1)\log{\left(1+\frac{A_\text{overlap}(D)q(h,H)}{AH-2\pi\left(\frac{D+\xi}{2}\right)^2q(h,H)}\right)}\nonumber\\
&\approx N_b(N+1)\frac{\xi^2\chi(\lambda_\text{br})q(h,H)}{AH}\nonumber\\
&=(N+1)\chi(\lambda_\text{br})\frac{q(h,H)}{H}\nonumber\\
&=\begin{cases} 
(N+1)\chi(\lambda_\text{br}),&\mbox{for } h>H \\ 
(N+1)\chi(\lambda_\text{br})\frac{h}{H},& \mbox{for } h<H, \end{cases}
\label{eqn:dF2}
\end{align}
where a large volume ($AH\gg 1$) was assumed for the brush system, with $A_\text{overlap}(D)=\xi^2\chi(\lambda_\text{br})$ and $\sigma\xi^2\simeq1$. 
Eq.~\eqref{eqn:dF2} suggests that $N$ and $\lambda_\text{br}$ (or $\sigma$) are the key parameters that determine the free energy gain upon the brush-induced clustering.

According to Eq.~\eqref{eqn:dF2} plotted against $N$ in Fig.~\ref{fig:height}B, the brush induced depletion interaction, quantified in terms of stability gain $-\beta\Delta F$ increases linearly with polymer length ($-\beta\Delta F\propto N$) when the brush is kept shorter than the height of the inclusion ($H<h$). 
However, as soon as the brush height exceeds the inclusion height ($H>h$), the free energy gain is reduced. 
When $H>h$, the same amount of accessible volume $A(H-h)$ is added regardless of the state of the two inclusions, increasing both the volume $V(D)$ and $V(r\geq D+\xi)$ accessible for brush polymers. 
This leads to the reduction of $-\beta\Delta F$. 
The factor $h/H$ that appears in the last line of Eq.\ref{eqn:dF2} quantifies the extent of this reduction in free energy gain (see Appendix B for further clarification).

For $H\gg h$, the free energy gain converges to 
\begin{align}
-\beta\Delta F\sim \frac{\chi(\lambda_\text{br})h}{\sigma^{1/3}b^{5/3}}< \chi(\lambda_\text{br})N,
\end{align}
where the inequality holds because of $h<H=N\sigma^{1/3}b^{5/3}$.   
Also, in the limit of $H\gg h$, it can be shown that $-\beta\Delta F\sim \sigma^{-1/12}h
$, which explains the $\sigma$-dependent limit of $\beta\Delta F$ at large $N$ in Fig.\ref{fig:height}B. 
The crossover point of polymer length $N^*$ changes with the grafting density as $N^\ast\simeq h\sigma^{-1/3}b^{-5/3}$. 

There is a crossover in the stability gain as well when the grafting density  ($\sigma$) is increased (Fig.~\ref{fig:height}C). 
The depletion free energy scales with $\sigma$ as 
\begin{align}
-\beta\Delta F\sim 
\begin{cases} 
(N+1)\sigma^{1/4},&\mbox{for } \sigma<\sigma^\ast \\ 
\frac{N+1}{N}\sigma^{-1/12},& \mbox{for } \sigma>\sigma^\ast,\
\end{cases}
\label{eqn:sigma}
\end{align}
with the crossover grafting density $\sigma^\ast b^2\simeq (h/Nb)^3$.  

\begin{figure}[t]
\centering
	\includegraphics[width=0.6\linewidth]{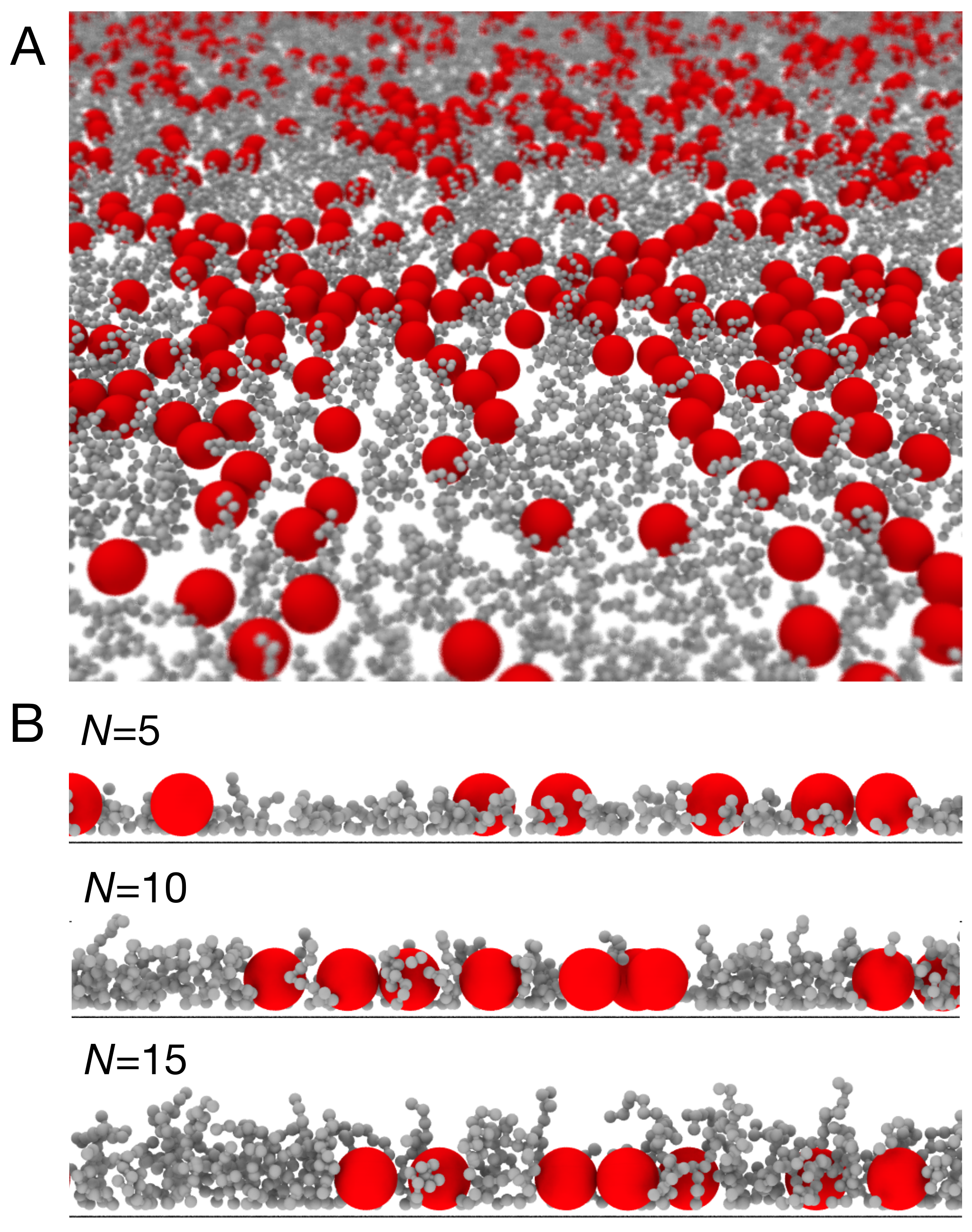}
	\caption{(A) A snapshot of simulations. 
	The spheres (red) and polymers (grey) represent membrane proteins and brush polymers grafted on the 2D surface, respectively.   
	(B) Lateral view of simulations for different brush sizes ($N=5$, 10, and 15). 
	 \label{fig:snapshot}}
\end{figure}

\section*{Numerical Results}

\subsection{Model}

%

The system is defined by $N_\text{b}$ brush polymers comprising the brush, 
and $M$ membrane proteins embedded in the brush on the 2D surface (Fig.~\ref{fig:snapshot}).  
The center of the protein, modeled as a sphere 
whose diameter (or vdW radius) is $D=5a$, is constrained on the surface at $z=D/2$, with a harmonic potential, to move only in parallel to the surface. 
The individual polymer consisting of $N$ segments (or $N+1$ monomers) is modeled using an energy potential for a bead-spring chain with self-avoidance. 
Each monomer with diameter $a$ is connected via the harmonic potential,
\begin{align}\label{eq:spring}
V_{\rm s}(r_{i,i+1})=\frac{k_\text{s}}{2}(r_{i,i+1}-b)^2,
\end{align}
where $k_\text{s}=3000~k_\text{B}T/a^2$ is the spring constant and $b=2^{1/6}a$ is the equilibrium bond length. 
Similarly to the protein, the first monomers of the chain, grafted to the surface at $z=a/2$, are free to move in the $xy$ plain, but constrained in the $z$ direction via a harmonic potential. 
Any non-grafted monomer whose distance from the grafting surface is $z\leq a$ is repelled by the Lennard-Jones (LJ) potential truncated at $z=a$,  
\begin{equation}
V_{\text{LJ}}^{\text{surf}}(z) =
\begin{cases}
4 k_\text{B} T \left[\left(\frac{a}{z}\right)^{12} -\left(\frac{a}{z}\right)^6 \right], & \text{ for $z \leq a$}\\
0,& \text{ for $z > a$}.
\end{cases} \label{eq:wall}      
\end{equation}
Both intra-chain and inter-chain monomer--monomer interactions as well as  protein--monomer and protein--protein interactions are modeled with LJ potential. 
\begin{align}
V_{\text{LJ}}^{\alpha\beta}(r_{ij})=
\begin{cases}
4\epsilon_{\alpha\beta}\left[\left(\frac{d_{\alpha\beta}}{r_{ij}}\right)^{12}-\left(\frac{d_{\alpha\beta}}{r_{ij}}\right)^6\right],&\text{ for $r_{ij}\leq r_c$}\\
0,&\text{ for $r_{ij}>r_c$}.
\end{cases}\label{eqn:ljpp}
\end{align}
Here, $\alpha$ and $\beta$ denote different particle types, $\alpha,\beta\in \{\text{m},\text{P}\}$, with m and P standing for monomer and protein. 
$r_{ij}$ is the distance between particles $i$ and $j$, $\epsilon_{\alpha \beta}$ is the strength of the interaction, and $d_{\alpha \beta}(=(d_\alpha+d_\beta)/2)$ is the contact distance between the particle types $\alpha$ and $\beta$. 
We have chosen $\beta\epsilon_{\alpha\beta}=1.0$ for all possible pairs of particle types; 
$d_\text{P}=5a$, $d_\text{m}=a$; 
$r_c=2.5\times d_\text{PP}$, $d_\text{mP}$, and $d_\text{mm}$ are the values of cut-off distance for protein-protein, monomer-protein, and monomer-monomer pairs, respectively. 
As a result, monomer--protein and monomer--monomer interactions are purely repulsive; and the protein--protein interactions in the absence of polymer brush are effectively under $\Theta$-solvent condition to yield a nearly vanishing second virial coefficient.

The simulation box has a dimension of $L_x =L_y=200a$ and $L_z=(N+1)b+\Delta$ with $\Delta=5a$, where $a$ is the basic length unit of our simulations. 
The system is periodic along the $x$ and $y$ directions and finite in the $z$ direction.  
With the fixed number of proteins $M=400$, 
the area fraction of the membrane proteins is $\phi_\text{P}=\pi(D/2)^2M/(L_x L_y)=0.2$, which corresponds to the surface density, $\sigma_\text{P} = 0.01/a^2$. 
The $\phi_\text{P}$ is related with $\sigma_\text{P}$ as 
$\phi_\text{P}=\sigma_\text{P}\times \pi (D/2)^2$. 
The grafting density of brush polymer is calculated using $\sigma = N_\text{b} /(L_x L_y - \pi (D/2)^2 M)$. In the simulations, $\sigma a^2$ is varied between 0.05 and 0.09. 

\subsection{Simulations} 
For the efficient sampling of the configurations of the polymer brush system including proteins, we used the low-friction Langevin dynamics to integrate the equation of motion \cite{VeitshansFoldDes97,Hyeon08JACS}. 
\begin{align}
m\ddot{x}_i = - \gamma \dot{x}_i -\partial_{x_i} V(\{\mathbf{r}_k\}) +\eta_i(t),
\label{eq:ld}
\end{align}
where $m$ is the mass of $i$-th particle. 
The characteristic time of the equation is set $\tau=  (m a^2/\epsilon)^{1/2}$ with the characteristic energy scale of inter-particle interaction $\epsilon=1k_\text{B}T$ specified in the energy potential $V(\{\mathbf{r}_k\})$. 
Then, the friction constant is set to $\gamma=0.05 m/\tau$. 
The last term $\eta_i(t)$ acting on the $i$-th particle ($i\in\{\text{m},\text{P}\}$) is the Gaussian white noise with zero mean, $\langle \eta_i(t)\rangle=0$, satisfying the fluctuation dissipation theorem, $\langle \eta_i(t)\eta_j(t ^\prime)\rangle=2\gamma k_\text{B}T\delta_{ij}\delta(t-t^{\prime})$. 
The equation of motion (Eq.~\eqref{eq:ld}) was integrated using the velocity-Verlet algorithm with the integration time step $\delta t = 0.0025\tau$ \cite{VeitshansFoldDes97,Hyeon08JACS}. 
After the pre-equilibration that fully randomizes the initial configurations of the system, the production runs of $4\times 10^8$ time steps were performed and collected for the statistical analysis. 

\begin{figure}[t]
	\centering
	\includegraphics[width=0.6\linewidth]{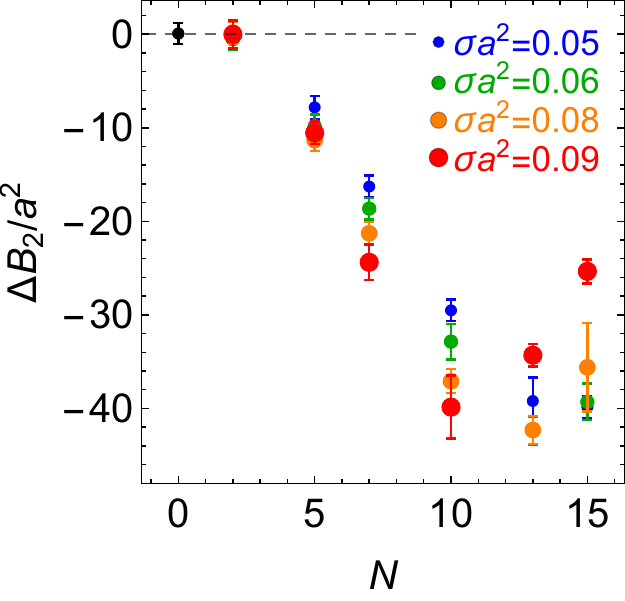}
	\caption{The measure of the brush polymer-induced protein--protein interaction, $\Delta B_2 =B_2-B_2^{\rm ref}$, as a function of the polymer brush size ($N$) for different grafting densities ($\sigma$). The data point at $N=0$ is for the protein-only reference system. 
	\label{fig:DB2N}}
\end{figure}

\subsection{Second virial coefficient}
The radial distribution function $g(r)$ between the membrane proteins (Fig.~S1) is associated with the second virial coefficient and is calculated for different set of parameters of brush size ($N$) and grafting density ($\sigma$) as follows. 
\begin{align}
B_2&=\frac{1}{2}\int (1-e^{-\beta u(\mathbf{r})})\text{d}\mathbf{r}\nonumber\\
&\simeq \pi\int_0^{\infty} (1-g(r))r\text{d}r.  
\end{align} 
We denote the second virial coefficient of a protein-only system as $B_2^\text{ref}$, and assess the depletion interaction in terms of $\Delta B_2=B_2-B_2^\text{ref}$, which can be related to the depletion induced free energy stabilization as $\beta\Delta F\sim \Delta B_2\sigma_\text{P}\sigma$.  
To simplify our interpretation of the simulation result, 
we have chosen the parameters for the protein--protein interaction to yield $B_2^\text{ref}\simeq 0$ (see Fig.~S2).

Overall trends of the simulation results indicate that 
the depletion interaction between the proteins increases with increasing grafting density ($\sigma$) and brush size ($N$); however, this trend is saturated or even inverted when the brush size is greater than a certain value (Fig.~\ref{fig:DB2N}).  
The non-monotonic dependence of the depletion interaction ($\Delta B_2$) on $N$ becomes more pronounced at high grafting density. 
Fig.~\ref{fig:DB2N} shows that the depletion effect for $\sigma a^2=0.09$ 
is maximized at $N=N^\ast\simeq 10$, at which the brush height ($H$) becomes comparable to the size of protein, ($D$).
This behavior is in agreement with 
the theoretical prediction of crossover at $h\simeq H=N^\ast\sigma^{1/3}b^{5/3}$ (Fig.~\ref{fig:height}B). 
With $h=5a$, $\sigma a^2=0.09$, and $b=2^{1/6}a$, we obtain $N^\ast=h\sigma^{-1/3}b^{-5/3}\simeq 9.2$ (see also Fig.~S3), which is in good agreement with Fig.~\ref{fig:DB2N}. 

\begin{figure*}[ht]
	\includegraphics[width=1.0\linewidth]{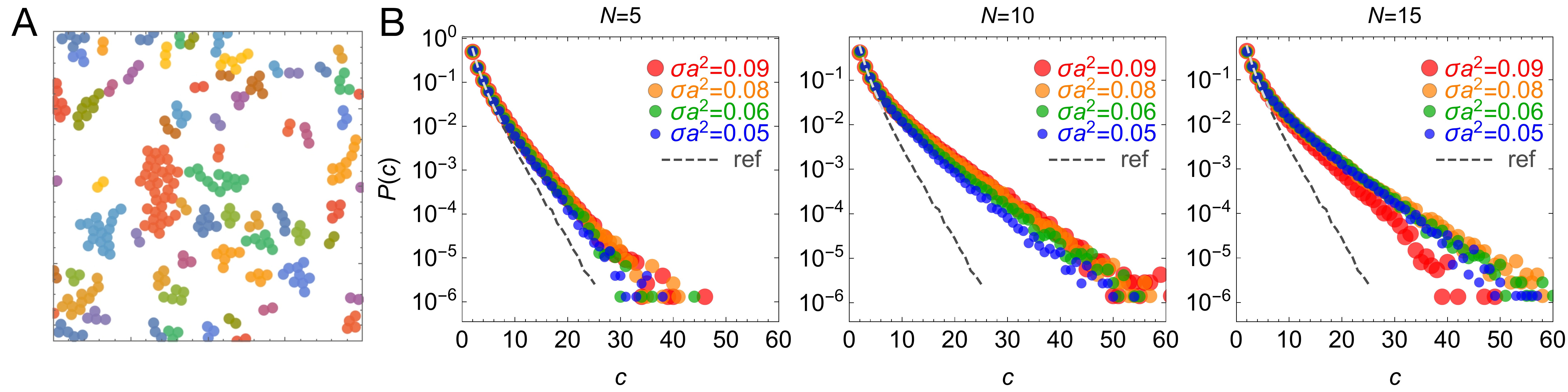}
	\caption{
	Cluster size distribution. 
	(A) A snapshot from simulation carried out with $N=10$, $\sigma_\text{P}a^2=0.01$, and $\sigma a^2=0.09$. 
	(B) The cluster size distribution, $P(c)$, with $\sigma_\text{P}a^2=0.01$ for varying brush sizes ($N=5$, 10, 15) and grafting densities ($\sigma$).  
	The dashed lines represent $P_\text{ref}(c)$, the cluster size distribution for the protein-only system. 
	\label{fig:Pm} 
	}
\end{figure*}

\subsection{Brush-induced protein clustering}

One of the goals of this study is to identify the condition that yields a large sized protein clustering.
To this end, we analyze the snapshots of simulations to calculate the cluster size distribution.  
We consider that two membrane proteins form a cluster of size two if the distance between them is less than the distance criterion of $6a$, which can be extended to identify a cluster of size $m$.

\begin{figure}[h!]
	\includegraphics[width=0.65\linewidth]{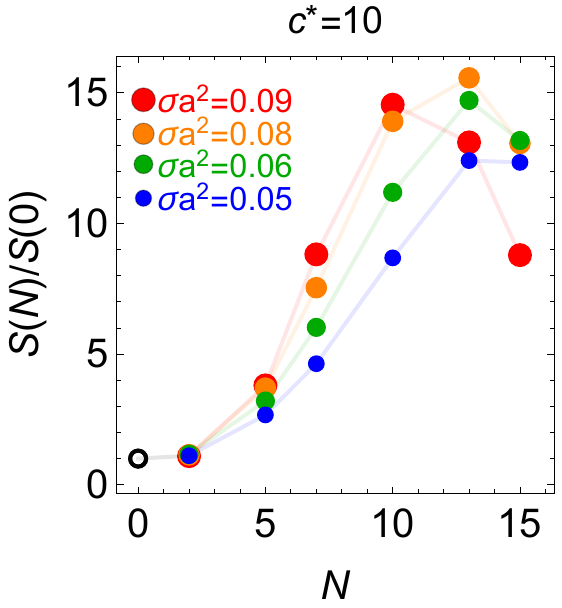}
	\caption{\label{fig:Signal}
	The intensity of signaling $S(N;\sigma)$ normalized by $S(N=0)$ (circle) is calculated based on Eq.~\eqref{eqn:S}, 
	as a function of the brush size ($N$) for different grafting densities ($\sigma$) with the threshold cluster size $c^\ast=10$.}
\end{figure}

Although the mean cluster size obtained from the simulation results is small ($\langle c\rangle\left[=\int_{c\geq 1} cP(c) \text{d} c\right] = 2-3$), $P(c)$s display long tails signifying the presence of large clusters (Fig.\ref{fig:Pm}).  
Deviation of $P(c)$ from that of the protein-only reference system ($P_\text{ref}(c)$) is observed at $c \gtrsim c^\ast \approx 10$ (Fig.\ref{fig:Pm}). 
With an assumption that the intensity of downstream signal ($S$) is proportional to the size of a cluster ($c>c^\ast$), which is greater than $c^\ast$, weighted by the population ($P(c)$),  
we evaluate the signal relayed from the protein clusters using 
\begin{align}
S(N,\sigma)\propto \int_{c \geq c^\ast}cP(c;N,\sigma) \text{d} c,
\label{eqn:S}
\end{align}
with $c^\ast=10$. 
The signal intensity calculated for varying grafting densities (Fig.~\ref{fig:Signal}) demonstrates a sigmoidal increase of $S$ as a function of brush size ($N$) up to $N\leq N^\ast$, beyond which $S$ decreases, suggestive of shrinking cluster size, reflecting the decrease of $|\Delta B_2|$. 
The mid-point of $S(N)$ shifts to a smaller $N$ from $N\simeq9$ to $N\simeq6$ as $\sigma$ increases from $\sigma a^2=0.05$ to 0.09.

\section*{Discussion}
The AO theory extended to the brush system (Eq.\ref{eqn:dF2}) differs from the hard sphere systems with two types (large and small spheres) in three dimensions (Eq.\ref{eqn:dF3}) in several aspects:  
(i) One of the key parameters $\lambda_\text{br}(=D/\xi)$ is the ratio of inclusion size ($D$) to blob size ($\xi$, grafting distance), 
whereas $\lambda(=R_L/R_S)$ is the ratio of large to small sphere sizes, $R_L$ and $R_S$.   
The blob size ($\xi\simeq bg^{\nu}$), equivalent to the grafting distance, is decided, independently from the size ($b$) of monomers, via the adaptation of polymer configuration. 
The term $\chi(\lambda_\text{br})$, which is a key determinant of the depletion free energy, is maximized for a larger $\lambda_\text{br}$ value under the condition of $H<h$;  
(ii) $|\beta\Delta F_\text{HS}|\sim \lambda$, whereas $|\beta\Delta F|\sim \sqrt{\lambda_\text{br}}$ for $\lambda_\text{br}\gg 1$;  
(iii) Whereas $\beta\Delta F_\text{HS}$, the depletion free energy of the hard sphere system, depends linearly on the volume fraction of crowders $\phi$ (Eq.\ref{eqn:dF3}), the dependence of area fraction of brush polymer (or grafting density, $\sigma$) is given as $\beta\Delta F\sim \lambda_{\text{br}}^{1/2}\sim\sigma^{1/4}$ for $\sigma<\sigma^\ast$ (Eq.\ref{eqn:sigma}). 
(iv) The non-monotonic dependence of depletion free energy on the brush size $N$ is unique to the brush-induced depletion interaction (see Appendix B); such feature is absent in the hard sphere systems in three dimensions.

\begin{figure}[b]
	\includegraphics[width=0.98\linewidth]{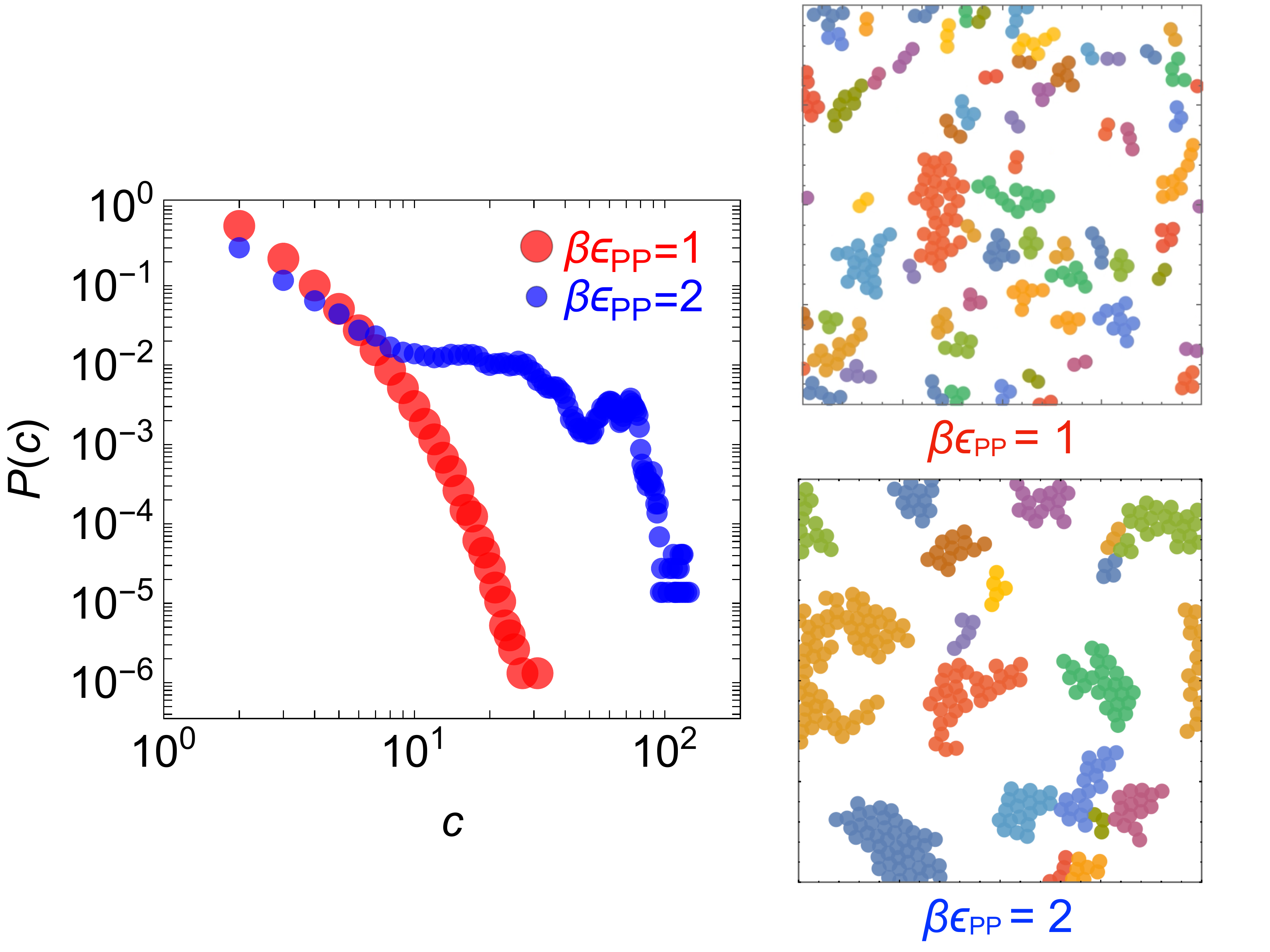}	
	\caption{The cluster size distribution, $P(c)$, for $\beta \epsilon_{\text{PP}}=1$ and 2. 
The two panels shown on the right are the snapshots of simulations at $\beta\epsilon_{\text{PP}}=1$ (top) and 2 (bottom). 
}
\label{fig:AmN}
\end{figure}

The general consensus on the protein clusters on cell surface is that 
the size of membrane protein assemblies is on the order of $\sim$ 100 nm \cite{lang2010Physiology,baumgart2016NatureMethods}.  
On the plasma membrane of T-cells, CD4 proteins form clusters of size varying from 50 to 300 nm \cite{lukevs2017NatComm}.   
The size of clusters formed by SNARE-protein syntaxin is 50 -- 60 nm, containing 50 -- 75 molecules \cite{sieber2006BJ}. 
Compared with the quantitative knowledge on nanodomains of membrane proteins, the size of protein clusters implicated in Fig.~\ref{fig:Pm}A is smaller. 
Besides the brush polymer enhanced assembly of protein cluster, 
one can consider other physical mechanisms that increase the effective attraction between proteins, 
such as inter-protein helix-helix interactions \cite{ben1996BJ,lorent2015structural,anbazhagan2010membrane}, protein sorting via hydrophobic mismatch \cite{schmidt2008PRL,milovanovic2015NatComm,west2009BJ}, membrane curvature \cite{mcmahon2005nature,reynwar2007nature}, and thermal Casimir-like long-range force resulting from membrane undulation \cite{goulian1993EPL,park1996JPI,Machta12PRL}.  
Upon increasing the LJ potential parameter from $\beta\epsilon_\text{PP}=1$ to $\beta\epsilon_\text{PP}=2$, which increases the direct protein--protein interaction drastically (Fig.~S2),   
the contribution of the tail part of $P(c)$ becomes significant, and a host of large and stable protein clusters are more frequently found (Fig.~\ref{fig:AmN}). 
For $\beta\epsilon_\text{PP}=2$, the protein cluster size could be as large as $m\approx 100$.  

\section*{Concluding Remarks} 
We have studied polymer brush-induced entropic force in a system of rigid bodies constrained to move on the surface. 
Both of our theory and simulation results show that the depletion free energy is non-monotonic function of brush height ($H$), which is determined by the brush size ($N$) and surface grafting density ($\sigma$). 
Our theoretical argument explaining the features of lateral depletion force is based on the AO theory, which takes only the volume accessible to individual brush polymers into consideration to calculate the depletion free energy in terms of geometrical factors ($N$ and $\lambda_\text{br}$), 
but ignores the effects of correlation between the brush polymers. 
Despite the simplicity of our theoretical argument, the main features of brush-induced depletion interaction observed in the simulation results are well captured. 

Our study confirms the depletion force induced assembly of protein clusters, although the size of protein domains is slightly smaller than that estimated from measurements. 
Given that the brush-induced depletion interaction considered here is merely one of many physical mechanisms of protein--protein attraction, 
of great significance is the semi-quantitative agreement with experimentally observed size of nano-domains. 
Our study reiterates that the entropic force, which is brush-induced, is of fundamental importance in cell membrane environment.  

\setcounter{subsection}{0}
\setcounter{equation}{0}
\renewcommand{\theequation}{A\arabic{equation}}%

\section*{Appendix}
\subsection{Depletion force}
The brush-induced 2D depletion force acting on the two objects is 
$\beta f(r)=-\left(\partial \beta F/\partial r\right)_\beta$, 
\begin{align}
\beta f(r)=\frac{N_\text{b}(N+1)A'_\text{overlap}(r)q(h,H)}{AH-\left[2\pi\left(\frac{D+\xi}{2}\right)^2-A_\text{overlap}(r)\right]q(h,H)}, 
\label{eqn:force}
\end{align}
for $D\leq r\leq D+\xi$ and $\beta f(r)=0$ for $r>D+\xi$. 
For very large system ($A\gg 2\pi\left((D+\xi)/2\right)^2$), 
the denominator of Eq.~\eqref{eqn:force} is dominated by the term $AH$, and the depletion force for $D\leq r\leq D+\xi$ simplifies to 
\begin{align}
\beta f(r)=-2\sigma (N+1)\left[\left(\frac{D+\xi}{2}\right)^2-\left(\frac{r}{2}\right)^2\right]^{1/2}\frac{q(h,H)}{H}, 
\end{align}
where the grafting density of polymer brush $\sigma= N_\text{b}/A$ was used. 
For $r>D+\xi$, $\beta f(r)=0$. 
It is noteworthy that the depletion force is always attractive ($f(r)<0$) for $D\leq r\leq D+\xi$. 

The free energy gain upon aggregation or 
the work needed to separate the two inclusions in the brush system apart beyond the distance $D+\xi$ is obtained by integrating the depletion force from $r=D$ to $r=D+\xi$, which yields the expression identical to Eq.\ref{eqn:dF2}.

\setcounter{equation}{0}
\renewcommand{\theequation}{B\arabic{equation}}%

\subsection{Non-monotonicity of depletion free energy gain with increasing brush polymer size ($H$)}
Here, we clarify how the non-monotonic change of $-\beta \Delta F$ arises with increasing $H$, starting from the expression of the free energy gain ($-\beta\Delta F$) given in the first line of Eq.\ref{eqn:dF2}. 
\begin{align}
-\beta\Delta F&\sim \frac{N_bH}{\sigma^{1/3}}\log{\frac{V(D)}{V(r\geq D+\xi)}}.
\end{align}
To begin, we define $a_c$ the area occupied by the inclusions when they are in contact, and $a$ the area occupied by the inclusions when they are separated beyond $r=D+\xi$. Other parameters $N_b$, $H$, $h$, and $A$ are already defined in the main text. 
Below we use the condition that the overlapping area $A_\text{overlap}(D)=a-a_c\equiv\delta a$ is small compared to $A$ ($\delta a/A\ll 1$). 
\begin{enumerate}[label=(\roman*)]
\item For $H<h$,
\begin{align}
-\beta\Delta F&\sim \frac{N_bH}{\sigma^{1/3}}\log{\frac{(A-a_c)h}{(A-a)h}}\nonumber\\
&=\frac{N_bH}{\sigma^{1/3}}\log{\left[1+\frac{\delta a}{A-a}\right]}\nonumber\\
&\approx \frac{N_b}{\sigma^{1/3}}\left(\frac{H}{1-a/A}\right)\frac{\delta a}{A}\nonumber\\
&\approx \frac{1}{\sigma^{1/3}}\left(\frac{H}{1-a/A}\right)\chi(\lambda_\text{br})
\end{align} 
where $\delta a=\xi^2\chi(\lambda_\text{br})$, $N_b/A=\sigma$, and $\sigma\xi^2\simeq 1$ were used to obtain the expression in the last line. 
Thus, for $H<h$, $-\beta\Delta F$ increase linearly with $H$. 

\item For $H\geq h$,
\begin{align}
-\beta\Delta F&\sim \frac{N_bH}{\sigma^{1/3}}\log{\frac{(AH-a_ch)}{(AH-ah)}}\nonumber\\
&= \frac{N_bH}{\sigma^{1/3}}\log{\left[1+\frac{\delta a\times h}{AH-ah}\right]}\nonumber\\
&\approx \frac{N_b}{\sigma^{1/3}}\left(\frac{h}{1-ah/AH}\right)\frac{\delta a}{A}\nonumber\\
&=\frac{1}{\sigma^{1/3}}\left(\frac{h}{1-ah/AH}\right)\chi(\lambda_\text{br})\end{align}
Thus,  for $H\geq h$, $-\beta \Delta F$ decreases with $H$ from 
$-\beta \Delta F=\frac{1}{\sigma^{1/3}}\left(\frac{h}{1-a/A}\right)\chi(\lambda_\text{br})$, which is the maximum value of $-\beta \Delta F$, and converges to $(h/\sigma^{1/3})\chi(\lambda_\text{br})$ when $H/h\gg 1$.  
\end{enumerate}

\section*{ACKNOWLEDGMENTS}
This study was supported by KIAS Individual Grants CG076001 (W.K.K.) and CG035003 (C.H.). We thank the Center for Advanced Computation in KIAS for providing computing resources.


\section*{References}

\begin{thebibliography}{76}
\expandafter\ifx\csname natexlab\endcsname\relax\def\natexlab#1{#1}\fi
\expandafter\ifx\csname bibnamefont\endcsname\relax
  \def\bibnamefont#1{#1}\fi
\expandafter\ifx\csname bibfnamefont\endcsname\relax
  \def\bibfnamefont#1{#1}\fi
\expandafter\ifx\csname citenamefont\endcsname\relax
  \def\citenamefont#1{#1}\fi
\expandafter\ifx\csname url\endcsname\relax
  \def\url#1{\texttt{#1}}\fi
\expandafter\ifx\csname urlprefix\endcsname\relax\def\urlprefix{URL }\fi
\providecommand{\bibinfo}[2]{#2}
\providecommand{\eprint}[2][]{\url{#2}}

\bibitem[{\citenamefont{Asakura and Oosawa}(1954)}]{asakura1954JCP}
\bibinfo{author}{\bibfnamefont{S.}~\bibnamefont{Asakura}} \bibnamefont{and}
  \bibinfo{author}{\bibfnamefont{F.}~\bibnamefont{Oosawa}},
  \bibinfo{journal}{J. Chem. Phys.} \textbf{\bibinfo{volume}{22}},
  \bibinfo{pages}{1255} (\bibinfo{year}{1954}).

\bibitem[{\citenamefont{Asakura and Oosawa}(1958)}]{Asakura58JPS}
\bibinfo{author}{\bibfnamefont{S.}~\bibnamefont{Asakura}} \bibnamefont{and}
  \bibinfo{author}{\bibfnamefont{F.}~\bibnamefont{Oosawa}},
  \bibinfo{journal}{J. Polym. Sci.} \textbf{\bibinfo{volume}{33}},
  \bibinfo{pages}{183} (\bibinfo{year}{1958}).

\bibitem[{\citenamefont{Marrenduzzo et~al.}(2006)\citenamefont{Marrenduzzo,
  Finan, and Cook}}]{Marrenduzzo06JCB}
\bibinfo{author}{\bibfnamefont{D.}~\bibnamefont{Marrenduzzo}},
  \bibinfo{author}{\bibfnamefont{K.}~\bibnamefont{Finan}}, \bibnamefont{and}
  \bibinfo{author}{\bibfnamefont{P.~R.} \bibnamefont{Cook}},
  \bibinfo{journal}{J. Cell. Biol.} \textbf{\bibinfo{volume}{175}},
  \bibinfo{pages}{681} (\bibinfo{year}{2006}).

\bibitem[{\citenamefont{Jeon et~al.}(2016)\citenamefont{Jeon, Hyeon, Jung, and
  Ha}}]{jeon2016Softmatter}
\bibinfo{author}{\bibfnamefont{C.}~\bibnamefont{Jeon}},
  \bibinfo{author}{\bibfnamefont{C.}~\bibnamefont{Hyeon}},
  \bibinfo{author}{\bibfnamefont{Y.}~\bibnamefont{Jung}}, \bibnamefont{and}
  \bibinfo{author}{\bibfnamefont{B.-Y.} \bibnamefont{Ha}},
  \bibinfo{journal}{Soft Matter} \textbf{\bibinfo{volume}{12}},
  \bibinfo{pages}{9786} (\bibinfo{year}{2016}).

\bibitem[{\citenamefont{Kang et~al.}(2015{\natexlab{a}})\citenamefont{Kang,
  Pincus, Hyeon, and Thirumalai}}]{Kang15PRL}
\bibinfo{author}{\bibfnamefont{H.}~\bibnamefont{Kang}},
  \bibinfo{author}{\bibfnamefont{P.~A.} \bibnamefont{Pincus}},
  \bibinfo{author}{\bibfnamefont{C.}~\bibnamefont{Hyeon}}, \bibnamefont{and}
  \bibinfo{author}{\bibfnamefont{D.}~\bibnamefont{Thirumalai}},
  \bibinfo{journal}{Phys. Rev. Lett.} \textbf{\bibinfo{volume}{114}},
  \bibinfo{pages}{068303} (\bibinfo{year}{2015}{\natexlab{a}}).

\bibitem[{\citenamefont{Phillips et~al.}(2009)\citenamefont{Phillips, Kondev,
  Theriot, Orme, and Garcia}}]{phillips2009physical}
\bibinfo{author}{\bibfnamefont{R.}~\bibnamefont{Phillips}},
  \bibinfo{author}{\bibfnamefont{J.}~\bibnamefont{Kondev}},
  \bibinfo{author}{\bibfnamefont{J.}~\bibnamefont{Theriot}},
  \bibinfo{author}{\bibfnamefont{N.}~\bibnamefont{Orme}}, \bibnamefont{and}
  \bibinfo{author}{\bibfnamefont{H.}~\bibnamefont{Garcia}},
  \emph{\bibinfo{title}{{Physical Biology of the Cell}}}
  (\bibinfo{publisher}{Garland Science New York}, \bibinfo{year}{2009}).

\bibitem[{\citenamefont{Roberts et~al.}(2011)\citenamefont{Roberts, Magis,
  Ortiz, Baumeister, and Luthey-Schulten}}]{Roberts2011PLOSCOMPBIO}
\bibinfo{author}{\bibfnamefont{E.}~\bibnamefont{Roberts}},
  \bibinfo{author}{\bibfnamefont{A.}~\bibnamefont{Magis}},
  \bibinfo{author}{\bibfnamefont{J.~O.} \bibnamefont{Ortiz}},
  \bibinfo{author}{\bibfnamefont{W.}~\bibnamefont{Baumeister}},
  \bibnamefont{and}
  \bibinfo{author}{\bibfnamefont{Z.}~\bibnamefont{Luthey-Schulten}},
  \bibinfo{journal}{PLoS Comput. Biol.} \textbf{\bibinfo{volume}{7}},
  \bibinfo{pages}{e1002010} (\bibinfo{year}{2011}).

\bibitem[{\citenamefont{Onsager}(1949)}]{Onsager49AnnNYAS}
\bibinfo{author}{\bibfnamefont{L.}~\bibnamefont{Onsager}},
  \bibinfo{journal}{Ann. NY Acad. Sci.} \textbf{\bibinfo{volume}{51}},
  \bibinfo{pages}{627} (\bibinfo{year}{1949}).

\bibitem[{\citenamefont{Dijkstra and Frenkel}(1994)}]{Dijkstra94PRL}
\bibinfo{author}{\bibfnamefont{M.}~\bibnamefont{Dijkstra}} \bibnamefont{and}
  \bibinfo{author}{\bibfnamefont{D.}~\bibnamefont{Frenkel}},
  \bibinfo{journal}{Phys. Rev. Lett.} \textbf{\bibinfo{volume}{72}},
  \bibinfo{pages}{298} (\bibinfo{year}{1994}).

\bibitem[{\citenamefont{Mao et~al.}(1995)\citenamefont{Mao, Cates, and
  Lekkerkerker}}]{mao1995PRL}
\bibinfo{author}{\bibfnamefont{Y.}~\bibnamefont{Mao}},
  \bibinfo{author}{\bibfnamefont{M.}~\bibnamefont{Cates}}, \bibnamefont{and}
  \bibinfo{author}{\bibfnamefont{H.}~\bibnamefont{Lekkerkerker}},
  \bibinfo{journal}{Phys. Rev. Lett.} \textbf{\bibinfo{volume}{75}},
  \bibinfo{pages}{4548} (\bibinfo{year}{1995}).

\bibitem[{\citenamefont{Mao et~al.}(1997)\citenamefont{Mao, Cates, and
  Lekkerkerker}}]{mao1997JCP}
\bibinfo{author}{\bibfnamefont{Y.}~\bibnamefont{Mao}},
  \bibinfo{author}{\bibfnamefont{M.}~\bibnamefont{Cates}}, \bibnamefont{and}
  \bibinfo{author}{\bibfnamefont{H.}~\bibnamefont{Lekkerkerker}},
  \bibinfo{journal}{J. Chem. Phys.} \textbf{\bibinfo{volume}{106}},
  \bibinfo{pages}{3721} (\bibinfo{year}{1997}).

\bibitem[{\citenamefont{Kang et~al.}(2015{\natexlab{b}})\citenamefont{Kang,
  Toan, Hyeon, and Thirumalai}}]{Kang15JACS}
\bibinfo{author}{\bibfnamefont{H.}~\bibnamefont{Kang}},
  \bibinfo{author}{\bibfnamefont{N.~M.} \bibnamefont{Toan}},
  \bibinfo{author}{\bibfnamefont{C.}~\bibnamefont{Hyeon}}, \bibnamefont{and}
  \bibinfo{author}{\bibfnamefont{D.}~\bibnamefont{Thirumalai}},
  \bibinfo{journal}{J. Am. Chem. Soc.} \textbf{\bibinfo{volume}{137}},
  \bibinfo{pages}{10970} (\bibinfo{year}{2015}{\natexlab{b}}).

\bibitem[{\citenamefont{Kaplan et~al.}(1994)\citenamefont{Kaplan, Rouke, Yodh,
  and Pine}}]{PhysRevLett.72.582}
\bibinfo{author}{\bibfnamefont{P.~D.} \bibnamefont{Kaplan}},
  \bibinfo{author}{\bibfnamefont{J.~L.} \bibnamefont{Rouke}},
  \bibinfo{author}{\bibfnamefont{A.~G.} \bibnamefont{Yodh}}, \bibnamefont{and}
  \bibinfo{author}{\bibfnamefont{D.~J.} \bibnamefont{Pine}},
  \bibinfo{journal}{Phys. Rev. Lett.} \textbf{\bibinfo{volume}{72}},
  \bibinfo{pages}{582} (\bibinfo{year}{1994}).

\bibitem[{\citenamefont{Dinsmore et~al.}(1998)\citenamefont{Dinsmore, Wong,
  Nelson, and Yodh}}]{Dinsmore98PRL}
\bibinfo{author}{\bibfnamefont{A.~D.} \bibnamefont{Dinsmore}},
  \bibinfo{author}{\bibfnamefont{D.~T.} \bibnamefont{Wong}},
  \bibinfo{author}{\bibfnamefont{P.}~\bibnamefont{Nelson}}, \bibnamefont{and}
  \bibinfo{author}{\bibfnamefont{A.~G.} \bibnamefont{Yodh}},
  \bibinfo{journal}{Phys. Rev. Lett.} \textbf{\bibinfo{volume}{80}},
  \bibinfo{pages}{409} (\bibinfo{year}{1998}).

\bibitem[{\citenamefont{Minton}(2000)}]{Minton00COSB}
\bibinfo{author}{\bibfnamefont{A.~P.} \bibnamefont{Minton}},
  \bibinfo{journal}{Curr. Opin. Struct. Biol.} \textbf{\bibinfo{volume}{10}},
  \bibinfo{pages}{34} (\bibinfo{year}{2000}).

\bibitem[{\citenamefont{McGuffee and Elcock}(2010)}]{mcguffee2010PLOS}
\bibinfo{author}{\bibfnamefont{S.~R.} \bibnamefont{McGuffee}} \bibnamefont{and}
  \bibinfo{author}{\bibfnamefont{A.~H.} \bibnamefont{Elcock}},
  \bibinfo{journal}{PLoS Comput Biol} \textbf{\bibinfo{volume}{6}},
  \bibinfo{pages}{e1000694} (\bibinfo{year}{2010}).

\bibitem[{\citenamefont{Ellis}(2001)}]{ellis2001TIBS}
\bibinfo{author}{\bibfnamefont{R.}~\bibnamefont{Ellis}},
  \bibinfo{journal}{Trends Biochem. Sci.} \textbf{\bibinfo{volume}{26}},
  \bibinfo{pages}{597} (\bibinfo{year}{2001}), ISSN \bibinfo{issn}{0968-0004}.

\bibitem[{\citenamefont{Sapir and Harries}(2015)}]{sapir2015COCI}
\bibinfo{author}{\bibfnamefont{L.}~\bibnamefont{Sapir}} \bibnamefont{and}
  \bibinfo{author}{\bibfnamefont{D.}~\bibnamefont{Harries}},
  \bibinfo{journal}{Current opinion in colloid \& interface science}
  \textbf{\bibinfo{volume}{20}}, \bibinfo{pages}{3} (\bibinfo{year}{2015}).

\bibitem[{\citenamefont{van~den Berg et~al.}(1999)\citenamefont{van~den Berg,
  Ellis, and Dobson}}]{van1999EMBO}
\bibinfo{author}{\bibfnamefont{B.}~\bibnamefont{van~den Berg}},
  \bibinfo{author}{\bibfnamefont{R.~J.} \bibnamefont{Ellis}}, \bibnamefont{and}
  \bibinfo{author}{\bibfnamefont{C.~M.} \bibnamefont{Dobson}},
  \bibinfo{journal}{EMBO J.} \textbf{\bibinfo{volume}{18}},
  \bibinfo{pages}{6927} (\bibinfo{year}{1999}).

\bibitem[{\citenamefont{Snir and Kamien}(2005)}]{Snir2005Science}
\bibinfo{author}{\bibfnamefont{Y.}~\bibnamefont{Snir}} \bibnamefont{and}
  \bibinfo{author}{\bibfnamefont{R.~D.} \bibnamefont{Kamien}},
  \bibinfo{journal}{Science} \textbf{\bibinfo{volume}{307}},
  \bibinfo{pages}{1067} (\bibinfo{year}{2005}).

\bibitem[{\citenamefont{Kudlay et~al.}(2009)\citenamefont{Kudlay, Cheung, and
  Thirumalai}}]{Kudlay09PRL}
\bibinfo{author}{\bibfnamefont{A.}~\bibnamefont{Kudlay}},
  \bibinfo{author}{\bibfnamefont{M.~S.} \bibnamefont{Cheung}},
  \bibnamefont{and}
  \bibinfo{author}{\bibfnamefont{D.}~\bibnamefont{Thirumalai}},
  \bibinfo{journal}{Phys. Rev. Lett.} \textbf{\bibinfo{volume}{102}},
  \bibinfo{pages}{118101} (\bibinfo{year}{2009}).

\bibitem[{\citenamefont{Zhou et~al.}(2008)\citenamefont{Zhou, Rivas, and
  Minton}}]{Zhou08ARB}
\bibinfo{author}{\bibfnamefont{H.~X.} \bibnamefont{Zhou}},
  \bibinfo{author}{\bibfnamefont{G.}~\bibnamefont{Rivas}}, \bibnamefont{and}
  \bibinfo{author}{\bibfnamefont{A.~P.} \bibnamefont{Minton}},
  \bibinfo{journal}{Annu. Rev. Biophys.} \textbf{\bibinfo{volume}{37}},
  \bibinfo{pages}{375} (\bibinfo{year}{2008}).

\bibitem[{\citenamefont{Elcock}(2010)}]{Elcock10COSB}
\bibinfo{author}{\bibfnamefont{A.}~\bibnamefont{Elcock}},
  \bibinfo{journal}{Curr. Opin. Struct. Biol.} \textbf{\bibinfo{volume}{20}},
  \bibinfo{pages}{196} (\bibinfo{year}{2010}).

\bibitem[{\citenamefont{Cheung et~al.}(2005)\citenamefont{Cheung, Klimov, and
  Thirumalai}}]{Cheung05PNAS}
\bibinfo{author}{\bibfnamefont{M.~S.} \bibnamefont{Cheung}},
  \bibinfo{author}{\bibfnamefont{D.}~\bibnamefont{Klimov}}, \bibnamefont{and}
  \bibinfo{author}{\bibfnamefont{D.}~\bibnamefont{Thirumalai}},
  \bibinfo{journal}{Proc. Natl. Acad. Sci. U. S. A.}
  \textbf{\bibinfo{volume}{102}}, \bibinfo{pages}{4753} (\bibinfo{year}{2005}).

\bibitem[{\citenamefont{Pincus et~al.}(2008)\citenamefont{Pincus, Hyeon, and
  Thirumalai}}]{Pincus08JACS}
\bibinfo{author}{\bibfnamefont{D.~L.} \bibnamefont{Pincus}},
  \bibinfo{author}{\bibfnamefont{C.}~\bibnamefont{Hyeon}}, \bibnamefont{and}
  \bibinfo{author}{\bibfnamefont{D.}~\bibnamefont{Thirumalai}},
  \bibinfo{journal}{J. Am. Chem. Soc.} \textbf{\bibinfo{volume}{130}},
  \bibinfo{pages}{7364} (\bibinfo{year}{2008}).

\bibitem[{\citenamefont{Kilburn et~al.}(2010)\citenamefont{Kilburn, Roh, Guo,
  Briber, and Woodson}}]{Kilburn10JACS}
\bibinfo{author}{\bibfnamefont{D.}~\bibnamefont{Kilburn}},
  \bibinfo{author}{\bibfnamefont{J.~H.} \bibnamefont{Roh}},
  \bibinfo{author}{\bibfnamefont{L.}~\bibnamefont{Guo}},
  \bibinfo{author}{\bibfnamefont{R.~M.} \bibnamefont{Briber}},
  \bibnamefont{and} \bibinfo{author}{\bibfnamefont{S.~A.}
  \bibnamefont{Woodson}}, \bibinfo{journal}{J. Am. Chem. Soc.}
  \textbf{\bibinfo{volume}{132}}, \bibinfo{pages}{8690} (\bibinfo{year}{2010}).

\bibitem[{\citenamefont{Denesyuk and Thirumalai}(2011)}]{Denesyuk11JACS}
\bibinfo{author}{\bibfnamefont{N.}~\bibnamefont{Denesyuk}} \bibnamefont{and}
  \bibinfo{author}{\bibfnamefont{D.}~\bibnamefont{Thirumalai}},
  \bibinfo{journal}{J. Am. Chem. Soc.} \textbf{\bibinfo{volume}{133}},
  \bibinfo{pages}{11858} (\bibinfo{year}{2011}).

\bibitem[{\citenamefont{Gasic et~al.}(2019)\citenamefont{Gasic, Boob,
  Prigozhin, Homouz, Daugherty, Gruebele, and Cheung}}]{PhysRevX.9.041035}
\bibinfo{author}{\bibfnamefont{A.~G.} \bibnamefont{Gasic}},
  \bibinfo{author}{\bibfnamefont{M.~M.} \bibnamefont{Boob}},
  \bibinfo{author}{\bibfnamefont{M.~B.} \bibnamefont{Prigozhin}},
  \bibinfo{author}{\bibfnamefont{D.}~\bibnamefont{Homouz}},
  \bibinfo{author}{\bibfnamefont{C.~M.} \bibnamefont{Daugherty}},
  \bibinfo{author}{\bibfnamefont{M.}~\bibnamefont{Gruebele}}, \bibnamefont{and}
  \bibinfo{author}{\bibfnamefont{M.~S.} \bibnamefont{Cheung}},
  \bibinfo{journal}{Phys. Rev. X} \textbf{\bibinfo{volume}{9}},
  \bibinfo{pages}{041035} (\bibinfo{year}{2019}).

\bibitem[{\citenamefont{Soranno et~al.}(2014)\citenamefont{Soranno, Koenig,
  Borgia, Hofmann, Zosel, Nettels, and Schuler}}]{Soranno2014PNAS}
\bibinfo{author}{\bibfnamefont{A.}~\bibnamefont{Soranno}},
  \bibinfo{author}{\bibfnamefont{I.}~\bibnamefont{Koenig}},
  \bibinfo{author}{\bibfnamefont{M.~B.} \bibnamefont{Borgia}},
  \bibinfo{author}{\bibfnamefont{H.}~\bibnamefont{Hofmann}},
  \bibinfo{author}{\bibfnamefont{F.}~\bibnamefont{Zosel}},
  \bibinfo{author}{\bibfnamefont{D.}~\bibnamefont{Nettels}}, \bibnamefont{and}
  \bibinfo{author}{\bibfnamefont{B.}~\bibnamefont{Schuler}},
  \bibinfo{journal}{Proc. Natl. Acad. Sci. U. S. A.}
  \textbf{\bibinfo{volume}{111}}, \bibinfo{pages}{4874} (\bibinfo{year}{2014}).

\bibitem[{\citenamefont{Li et~al.}(2009)\citenamefont{Li, Berg, and
  Elf}}]{Li09NaturePhysics}
\bibinfo{author}{\bibfnamefont{G.-W.} \bibnamefont{Li}},
  \bibinfo{author}{\bibfnamefont{O.~G.} \bibnamefont{Berg}}, \bibnamefont{and}
  \bibinfo{author}{\bibfnamefont{J.}~\bibnamefont{Elf}},
  \bibinfo{journal}{Nature Phys.} \textbf{\bibinfo{volume}{5}},
  \bibinfo{pages}{294 } (\bibinfo{year}{2009}).

\bibitem[{\citenamefont{Kim et~al.}(2011)\citenamefont{Kim, Backman, and
  Szleifer}}]{Kim11PRL}
\bibinfo{author}{\bibfnamefont{J.~S.} \bibnamefont{Kim}},
  \bibinfo{author}{\bibfnamefont{V.}~\bibnamefont{Backman}}, \bibnamefont{and}
  \bibinfo{author}{\bibfnamefont{I.}~\bibnamefont{Szleifer}},
  \bibinfo{journal}{Phys. Rev. Lett.} \textbf{\bibinfo{volume}{106}},
  \bibinfo{pages}{168102} (\bibinfo{year}{2011}).

\bibitem[{\citenamefont{Brackley et~al.}(2013)\citenamefont{Brackley, Cates,
  and Marenduzzo}}]{Brackley13PRL}
\bibinfo{author}{\bibfnamefont{C.~A.} \bibnamefont{Brackley}},
  \bibinfo{author}{\bibfnamefont{M.~E.} \bibnamefont{Cates}}, \bibnamefont{and}
  \bibinfo{author}{\bibfnamefont{D.}~\bibnamefont{Marenduzzo}},
  \bibinfo{journal}{Phys. Rev. Lett.} \textbf{\bibinfo{volume}{111}},
  \bibinfo{pages}{108101} (\bibinfo{year}{2013}).

\bibitem[{\citenamefont{Reddy et~al.}(2017)\citenamefont{Reddy, Tripathy,
  Vershinin, Tanenbaum, Xu, Mattson-Hoss, Arabi, Chapman, Doolin, Hyeon
  et~al.}}]{Reddy2017Traffic}
\bibinfo{author}{\bibfnamefont{B.~J.} \bibnamefont{Reddy}},
  \bibinfo{author}{\bibfnamefont{S.}~\bibnamefont{Tripathy}},
  \bibinfo{author}{\bibfnamefont{M.}~\bibnamefont{Vershinin}},
  \bibinfo{author}{\bibfnamefont{M.~E.} \bibnamefont{Tanenbaum}},
  \bibinfo{author}{\bibfnamefont{J.}~\bibnamefont{Xu}},
  \bibinfo{author}{\bibfnamefont{M.}~\bibnamefont{Mattson-Hoss}},
  \bibinfo{author}{\bibfnamefont{K.}~\bibnamefont{Arabi}},
  \bibinfo{author}{\bibfnamefont{D.}~\bibnamefont{Chapman}},
  \bibinfo{author}{\bibfnamefont{T.}~\bibnamefont{Doolin}},
  \bibinfo{author}{\bibfnamefont{C.}~\bibnamefont{Hyeon}},
  \bibnamefont{et~al.}, \bibinfo{journal}{Traffic}
  \textbf{\bibinfo{volume}{18}}, \bibinfo{pages}{658} (\bibinfo{year}{2017}).

\bibitem[{\citenamefont{Nettesheim et~al.}(2020)\citenamefont{Nettesheim,
  Nabti, Murade, Jaffe, King, and Shubeita}}]{nettesheim2020NatPhys}
\bibinfo{author}{\bibfnamefont{G.}~\bibnamefont{Nettesheim}},
  \bibinfo{author}{\bibfnamefont{I.}~\bibnamefont{Nabti}},
  \bibinfo{author}{\bibfnamefont{C.~U.} \bibnamefont{Murade}},
  \bibinfo{author}{\bibfnamefont{G.~R.} \bibnamefont{Jaffe}},
  \bibinfo{author}{\bibfnamefont{S.~J.} \bibnamefont{King}}, \bibnamefont{and}
  \bibinfo{author}{\bibfnamefont{G.~T.} \bibnamefont{Shubeita}},
  \bibinfo{journal}{Nature Physics} \textbf{\bibinfo{volume}{16}},
  \bibinfo{pages}{1144} (\bibinfo{year}{2020}).

\bibitem[{\citenamefont{Dzubiella et~al.}(2003)\citenamefont{Dzubiella,
  L{\"o}wen, and Likos}}]{dzubiella2003depletion}
\bibinfo{author}{\bibfnamefont{J.}~\bibnamefont{Dzubiella}},
  \bibinfo{author}{\bibfnamefont{H.}~\bibnamefont{L{\"o}wen}},
  \bibnamefont{and} \bibinfo{author}{\bibfnamefont{C.}~\bibnamefont{Likos}},
  \bibinfo{journal}{Phys. Rev. Lett.} \textbf{\bibinfo{volume}{91}},
  \bibinfo{pages}{248301} (\bibinfo{year}{2003}).

\bibitem[{\citenamefont{Angelani et~al.}(2011)\citenamefont{Angelani, Maggi,
  Bernardini, Rizzo, and Di~Leonardo}}]{angelani2011PRL}
\bibinfo{author}{\bibfnamefont{L.}~\bibnamefont{Angelani}},
  \bibinfo{author}{\bibfnamefont{C.}~\bibnamefont{Maggi}},
  \bibinfo{author}{\bibfnamefont{M.}~\bibnamefont{Bernardini}},
  \bibinfo{author}{\bibfnamefont{A.}~\bibnamefont{Rizzo}}, \bibnamefont{and}
  \bibinfo{author}{\bibfnamefont{R.}~\bibnamefont{Di~Leonardo}},
  \bibinfo{journal}{Phys. Rev. Lett.} \textbf{\bibinfo{volume}{107}},
  \bibinfo{pages}{138302} (\bibinfo{year}{2011}).

\bibitem[{\citenamefont{Sanchez et~al.}(2012)\citenamefont{Sanchez, Chen,
  DeCamp, Heymann, and Dogic}}]{sanchez2012Nature}
\bibinfo{author}{\bibfnamefont{T.}~\bibnamefont{Sanchez}},
  \bibinfo{author}{\bibfnamefont{D.~T.} \bibnamefont{Chen}},
  \bibinfo{author}{\bibfnamefont{S.~J.} \bibnamefont{DeCamp}},
  \bibinfo{author}{\bibfnamefont{M.}~\bibnamefont{Heymann}}, \bibnamefont{and}
  \bibinfo{author}{\bibfnamefont{Z.}~\bibnamefont{Dogic}},
  \bibinfo{journal}{Nature} \textbf{\bibinfo{volume}{491}},
  \bibinfo{pages}{431} (\bibinfo{year}{2012}).

\bibitem[{\citenamefont{Ni et~al.}(2015)\citenamefont{Ni, Stuart, and
  Bolhuis}}]{ni2015PRL}
\bibinfo{author}{\bibfnamefont{R.}~\bibnamefont{Ni}},
  \bibinfo{author}{\bibfnamefont{M.~A.~C.} \bibnamefont{Stuart}},
  \bibnamefont{and} \bibinfo{author}{\bibfnamefont{P.~G.}
  \bibnamefont{Bolhuis}}, \bibinfo{journal}{Phys. Rev. Lett.}
  \textbf{\bibinfo{volume}{114}}, \bibinfo{pages}{018302}
  (\bibinfo{year}{2015}).

\bibitem[{\citenamefont{Huber et~al.}(2018)\citenamefont{Huber, Suzuki,
  Kr{\"u}ger, Frey, and Bausch}}]{huber2018science}
\bibinfo{author}{\bibfnamefont{L.}~\bibnamefont{Huber}},
  \bibinfo{author}{\bibfnamefont{R.}~\bibnamefont{Suzuki}},
  \bibinfo{author}{\bibfnamefont{T.}~\bibnamefont{Kr{\"u}ger}},
  \bibinfo{author}{\bibfnamefont{E.}~\bibnamefont{Frey}}, \bibnamefont{and}
  \bibinfo{author}{\bibfnamefont{A.}~\bibnamefont{Bausch}},
  \bibinfo{journal}{Science} \textbf{\bibinfo{volume}{361}},
  \bibinfo{pages}{255} (\bibinfo{year}{2018}).

\bibitem[{\citenamefont{Sintes and Baumg{\"a}rtner}(1997)}]{sintes1997BJ}
\bibinfo{author}{\bibfnamefont{T.}~\bibnamefont{Sintes}} \bibnamefont{and}
  \bibinfo{author}{\bibfnamefont{A.}~\bibnamefont{Baumg{\"a}rtner}},
  \bibinfo{journal}{Biophys. J.} \textbf{\bibinfo{volume}{73}},
  \bibinfo{pages}{2251} (\bibinfo{year}{1997}).

\bibitem[{\citenamefont{Suda et~al.}(2020)\citenamefont{Suda, Suematsu, and
  Akiyama}}]{suda2020lateral}
\bibinfo{author}{\bibfnamefont{K.}~\bibnamefont{Suda}},
  \bibinfo{author}{\bibfnamefont{A.}~\bibnamefont{Suematsu}}, \bibnamefont{and}
  \bibinfo{author}{\bibfnamefont{R.}~\bibnamefont{Akiyama}},
  \bibinfo{journal}{arXiv preprint arXiv:2011.06232}  (\bibinfo{year}{2020}).

\bibitem[{\citenamefont{Paszek et~al.}(2014)\citenamefont{Paszek, DuFort,
  Rossier, Bainer, Mouw, Godula, Hudak, Lakins, Wijekoon, Cassereau
  et~al.}}]{Paszak14Nature}
\bibinfo{author}{\bibfnamefont{M.~J.} \bibnamefont{Paszek}},
  \bibinfo{author}{\bibfnamefont{C.~C.} \bibnamefont{DuFort}},
  \bibinfo{author}{\bibfnamefont{O.}~\bibnamefont{Rossier}},
  \bibinfo{author}{\bibfnamefont{R.}~\bibnamefont{Bainer}},
  \bibinfo{author}{\bibfnamefont{J.~K.} \bibnamefont{Mouw}},
  \bibinfo{author}{\bibfnamefont{K.}~\bibnamefont{Godula}},
  \bibinfo{author}{\bibfnamefont{J.~E.} \bibnamefont{Hudak}},
  \bibinfo{author}{\bibfnamefont{J.~N.} \bibnamefont{Lakins}},
  \bibinfo{author}{\bibfnamefont{A.~C.} \bibnamefont{Wijekoon}},
  \bibinfo{author}{\bibfnamefont{L.}~\bibnamefont{Cassereau}},
  \bibnamefont{et~al.}, \bibinfo{journal}{Nature}
  \textbf{\bibinfo{volume}{511}}, \bibinfo{pages}{319} (\bibinfo{year}{2014}).

\bibitem[{\citenamefont{Bakker et~al.}(2012)\citenamefont{Bakker, Eich,
  Torreno-Pina, Diez-Ahedo, Perez-Samper, van Zanten, Figdor, Cambi, and
  Garcia-Parajo}}]{bakker2012PNAS}
\bibinfo{author}{\bibfnamefont{G.~J.} \bibnamefont{Bakker}},
  \bibinfo{author}{\bibfnamefont{C.}~\bibnamefont{Eich}},
  \bibinfo{author}{\bibfnamefont{J.~A.} \bibnamefont{Torreno-Pina}},
  \bibinfo{author}{\bibfnamefont{R.}~\bibnamefont{Diez-Ahedo}},
  \bibinfo{author}{\bibfnamefont{G.}~\bibnamefont{Perez-Samper}},
  \bibinfo{author}{\bibfnamefont{T.~S.} \bibnamefont{van Zanten}},
  \bibinfo{author}{\bibfnamefont{C.~G.} \bibnamefont{Figdor}},
  \bibinfo{author}{\bibfnamefont{A.}~\bibnamefont{Cambi}}, \bibnamefont{and}
  \bibinfo{author}{\bibfnamefont{M.~F.} \bibnamefont{Garcia-Parajo}},
  \bibinfo{journal}{Proc. Natl. Acad. Sci. U. S. A.}
  \textbf{\bibinfo{volume}{109}}, \bibinfo{pages}{4869} (\bibinfo{year}{2012}).

\bibitem[{\citenamefont{Selhuber-Unkel
  et~al.}(2008)\citenamefont{Selhuber-Unkel, L{\'o}pez-Garc{\'\i}a, Kessler,
  and Spatz}}]{selhuber2008BJ}
\bibinfo{author}{\bibfnamefont{C.}~\bibnamefont{Selhuber-Unkel}},
  \bibinfo{author}{\bibfnamefont{M.}~\bibnamefont{L{\'o}pez-Garc{\'\i}a}},
  \bibinfo{author}{\bibfnamefont{H.}~\bibnamefont{Kessler}}, \bibnamefont{and}
  \bibinfo{author}{\bibfnamefont{J.~P.} \bibnamefont{Spatz}},
  \bibinfo{journal}{Biophys. J.} \textbf{\bibinfo{volume}{95}},
  \bibinfo{pages}{5424} (\bibinfo{year}{2008}).

\bibitem[{\citenamefont{van Zanten et~al.}(2010)\citenamefont{van Zanten,
  G{\'o}mez, Manzo, Cambi, Buceta, Reigada, and Garcia-Parajo}}]{van2010PNAS}
\bibinfo{author}{\bibfnamefont{T.~S.} \bibnamefont{van Zanten}},
  \bibinfo{author}{\bibfnamefont{J.}~\bibnamefont{G{\'o}mez}},
  \bibinfo{author}{\bibfnamefont{C.}~\bibnamefont{Manzo}},
  \bibinfo{author}{\bibfnamefont{A.}~\bibnamefont{Cambi}},
  \bibinfo{author}{\bibfnamefont{J.}~\bibnamefont{Buceta}},
  \bibinfo{author}{\bibfnamefont{R.}~\bibnamefont{Reigada}}, \bibnamefont{and}
  \bibinfo{author}{\bibfnamefont{M.~F.} \bibnamefont{Garcia-Parajo}},
  \bibinfo{journal}{Proc. Natl. Acad. Sci. U. S. A.}
  \textbf{\bibinfo{volume}{107}}, \bibinfo{pages}{15437}
  (\bibinfo{year}{2010}).

\bibitem[{\citenamefont{Garcia-Parajo et~al.}(2014)\citenamefont{Garcia-Parajo,
  Cambi, Torreno-Pina, Thompson, and Jacobson}}]{garcia2014JCS}
\bibinfo{author}{\bibfnamefont{M.~F.} \bibnamefont{Garcia-Parajo}},
  \bibinfo{author}{\bibfnamefont{A.}~\bibnamefont{Cambi}},
  \bibinfo{author}{\bibfnamefont{J.~A.} \bibnamefont{Torreno-Pina}},
  \bibinfo{author}{\bibfnamefont{N.}~\bibnamefont{Thompson}}, \bibnamefont{and}
  \bibinfo{author}{\bibfnamefont{K.}~\bibnamefont{Jacobson}},
  \bibinfo{journal}{J. Cell. Sci.} \textbf{\bibinfo{volume}{127}},
  \bibinfo{pages}{4995} (\bibinfo{year}{2014}).

\bibitem[{\citenamefont{Soubias et~al.}(2015)\citenamefont{Soubias, Teague~Jr,
  Hines, and Gawrisch}}]{soubias2015BJ}
\bibinfo{author}{\bibfnamefont{O.}~\bibnamefont{Soubias}},
  \bibinfo{author}{\bibfnamefont{W.~E.} \bibnamefont{Teague~Jr}},
  \bibinfo{author}{\bibfnamefont{K.~G.} \bibnamefont{Hines}}, \bibnamefont{and}
  \bibinfo{author}{\bibfnamefont{K.}~\bibnamefont{Gawrisch}},
  \bibinfo{journal}{Biophys. J.} \textbf{\bibinfo{volume}{108}},
  \bibinfo{pages}{1125} (\bibinfo{year}{2015}).

\bibitem[{\citenamefont{Kusumi and Hyde}(1982)}]{kusumi1982Biochem}
\bibinfo{author}{\bibfnamefont{A.}~\bibnamefont{Kusumi}} \bibnamefont{and}
  \bibinfo{author}{\bibfnamefont{J.~S.} \bibnamefont{Hyde}},
  \bibinfo{journal}{Biochemistry} \textbf{\bibinfo{volume}{21}},
  \bibinfo{pages}{5978} (\bibinfo{year}{1982}).

\bibitem[{\citenamefont{Kim and O'Shaughnessy}(2006)}]{kim2006macromolecules}
\bibinfo{author}{\bibfnamefont{J.~U.} \bibnamefont{Kim}} \bibnamefont{and}
  \bibinfo{author}{\bibfnamefont{B.}~\bibnamefont{O'Shaughnessy}},
  \bibinfo{journal}{Macromolecules} \textbf{\bibinfo{volume}{39}},
  \bibinfo{pages}{413} (\bibinfo{year}{2006}).

\bibitem[{\citenamefont{Spencer and Ha}(2021)}]{spencer2021macromolecules}
\bibinfo{author}{\bibfnamefont{R.~K.} \bibnamefont{Spencer}} \bibnamefont{and}
  \bibinfo{author}{\bibfnamefont{B.-Y.} \bibnamefont{Ha}},
  \bibinfo{journal}{Macromolecules} \textbf{\bibinfo{volume}{54}},
  \bibinfo{pages}{1304–1313} (\bibinfo{year}{2021}).

\bibitem[{\citenamefont{Williams et~al.}(1994)\citenamefont{Williams, Hughes,
  O'Toole, and Ginsberg}}]{williams1994TCB}
\bibinfo{author}{\bibfnamefont{M.~J.} \bibnamefont{Williams}},
  \bibinfo{author}{\bibfnamefont{P.~E.} \bibnamefont{Hughes}},
  \bibinfo{author}{\bibfnamefont{T.~E.} \bibnamefont{O'Toole}},
  \bibnamefont{and} \bibinfo{author}{\bibfnamefont{M.~H.}
  \bibnamefont{Ginsberg}}, \bibinfo{journal}{Trends in cell biology}
  \textbf{\bibinfo{volume}{4}}, \bibinfo{pages}{109} (\bibinfo{year}{1994}).

\bibitem[{\citenamefont{Kornberg et~al.}(1992)\citenamefont{Kornberg, Earp,
  Parsons, Schaller, and Juliano}}]{kornberg1992JBC}
\bibinfo{author}{\bibfnamefont{L.}~\bibnamefont{Kornberg}},
  \bibinfo{author}{\bibfnamefont{H.~S.} \bibnamefont{Earp}},
  \bibinfo{author}{\bibfnamefont{J.~T.} \bibnamefont{Parsons}},
  \bibinfo{author}{\bibfnamefont{M.}~\bibnamefont{Schaller}}, \bibnamefont{and}
  \bibinfo{author}{\bibfnamefont{R.}~\bibnamefont{Juliano}},
  \bibinfo{journal}{J. Biol. Chem.} \textbf{\bibinfo{volume}{267}},
  \bibinfo{pages}{23439} (\bibinfo{year}{1992}).

\bibitem[{\citenamefont{Paszek et~al.}(2009)\citenamefont{Paszek, Boettiger,
  Weaver, and Hammer}}]{paszek2009PLOS}
\bibinfo{author}{\bibfnamefont{M.~J.} \bibnamefont{Paszek}},
  \bibinfo{author}{\bibfnamefont{D.}~\bibnamefont{Boettiger}},
  \bibinfo{author}{\bibfnamefont{V.~M.} \bibnamefont{Weaver}},
  \bibnamefont{and} \bibinfo{author}{\bibfnamefont{D.~A.}
  \bibnamefont{Hammer}}, \bibinfo{journal}{PLoS Comput Biol}
  \textbf{\bibinfo{volume}{5}}, \bibinfo{pages}{e1000604}
  (\bibinfo{year}{2009}).

\bibitem[{\citenamefont{Cheng et~al.}(2020)\citenamefont{Cheng, Wan, Huang, Li,
  Genin, Mofrad, Lu, Xu, and Lin}}]{cheng2020SciAdv}
\bibinfo{author}{\bibfnamefont{B.}~\bibnamefont{Cheng}},
  \bibinfo{author}{\bibfnamefont{W.}~\bibnamefont{Wan}},
  \bibinfo{author}{\bibfnamefont{G.}~\bibnamefont{Huang}},
  \bibinfo{author}{\bibfnamefont{Y.}~\bibnamefont{Li}},
  \bibinfo{author}{\bibfnamefont{G.~M.} \bibnamefont{Genin}},
  \bibinfo{author}{\bibfnamefont{M.~R.} \bibnamefont{Mofrad}},
  \bibinfo{author}{\bibfnamefont{T.~J.} \bibnamefont{Lu}},
  \bibinfo{author}{\bibfnamefont{F.}~\bibnamefont{Xu}}, \bibnamefont{and}
  \bibinfo{author}{\bibfnamefont{M.}~\bibnamefont{Lin}}, \bibinfo{journal}{Sci.
  Adv.} \textbf{\bibinfo{volume}{6}}, \bibinfo{pages}{eaax1909}
  (\bibinfo{year}{2020}).

\bibitem[{\citenamefont{de~Gennes}(1980)}]{deGennes1980Macromolecules}
\bibinfo{author}{\bibfnamefont{P.~G.} \bibnamefont{de~Gennes}},
  \bibinfo{journal}{Macromolecules} \textbf{\bibinfo{volume}{13}},
  \bibinfo{pages}{1069} (\bibinfo{year}{1980}).

\bibitem[{\citenamefont{Liu et~al.}(2017)\citenamefont{Liu, Pincus, and
  Hyeon}}]{liu2017Macromolecules}
\bibinfo{author}{\bibfnamefont{L.}~\bibnamefont{Liu}},
  \bibinfo{author}{\bibfnamefont{P.~A.} \bibnamefont{Pincus}},
  \bibnamefont{and} \bibinfo{author}{\bibfnamefont{C.}~\bibnamefont{Hyeon}},
  \bibinfo{journal}{Macromolecules} \textbf{\bibinfo{volume}{50}},
  \bibinfo{pages}{1579} (\bibinfo{year}{2017}).

\bibitem[{\citenamefont{Liu and Hyeon}(2018)}]{liu2018JCP}
\bibinfo{author}{\bibfnamefont{L.}~\bibnamefont{Liu}} \bibnamefont{and}
  \bibinfo{author}{\bibfnamefont{C.}~\bibnamefont{Hyeon}}, \bibinfo{journal}{J.
  Chem. Phys.} \textbf{\bibinfo{volume}{149}}, \bibinfo{pages}{163302}
  (\bibinfo{year}{2018}).

\bibitem[{\citenamefont{Alexander}(1977)}]{Alexander77JP}
\bibinfo{author}{\bibfnamefont{S.}~\bibnamefont{Alexander}},
  \bibinfo{journal}{J. Phys.} \textbf{\bibinfo{volume}{38}},
  \bibinfo{pages}{983} (\bibinfo{year}{1977}).

\bibitem[{\citenamefont{Rubinstein et~al.}(2003)\citenamefont{Rubinstein, Colby
  et~al.}}]{rubinstein2003polymer}
\bibinfo{author}{\bibfnamefont{M.}~\bibnamefont{Rubinstein}},
  \bibinfo{author}{\bibfnamefont{R.~H.} \bibnamefont{Colby}},
  \bibnamefont{et~al.}, \emph{\bibinfo{title}{Polymer physics}},
  vol.~\bibinfo{volume}{23} (\bibinfo{publisher}{Oxford university press New
  York}, \bibinfo{year}{2003}).

\bibitem[{\citenamefont{Veitshans et~al.}(1997)\citenamefont{Veitshans, Klimov,
  and Thirumalai}}]{VeitshansFoldDes97}
\bibinfo{author}{\bibfnamefont{T.}~\bibnamefont{Veitshans}},
  \bibinfo{author}{\bibfnamefont{D.}~\bibnamefont{Klimov}}, \bibnamefont{and}
  \bibinfo{author}{\bibfnamefont{D.}~\bibnamefont{Thirumalai}},
  \bibinfo{journal}{Folding Des.} \textbf{\bibinfo{volume}{2}},
  \bibinfo{pages}{1} (\bibinfo{year}{1997}).

\bibitem[{\citenamefont{Hyeon and Thirumalai}(2008)}]{Hyeon08JACS}
\bibinfo{author}{\bibfnamefont{C.}~\bibnamefont{Hyeon}} \bibnamefont{and}
  \bibinfo{author}{\bibfnamefont{D.}~\bibnamefont{Thirumalai}},
  \bibinfo{journal}{J. Am. Chem. Soc.} \textbf{\bibinfo{volume}{130}},
  \bibinfo{pages}{1538} (\bibinfo{year}{2008}).

\bibitem[{\citenamefont{Lang and Rizzoli}(2010)}]{lang2010Physiology}
\bibinfo{author}{\bibfnamefont{T.}~\bibnamefont{Lang}} \bibnamefont{and}
  \bibinfo{author}{\bibfnamefont{S.~O.} \bibnamefont{Rizzoli}},
  \bibinfo{journal}{Physiology} \textbf{\bibinfo{volume}{25}},
  \bibinfo{pages}{116} (\bibinfo{year}{2010}).

\bibitem[{\citenamefont{Baumgart et~al.}(2016)\citenamefont{Baumgart, Arnold,
  Leskovar, Staszek, F{\"o}lser, Weghuber, Stockinger, and
  Sch{\"u}tz}}]{baumgart2016NatureMethods}
\bibinfo{author}{\bibfnamefont{F.}~\bibnamefont{Baumgart}},
  \bibinfo{author}{\bibfnamefont{A.~M.} \bibnamefont{Arnold}},
  \bibinfo{author}{\bibfnamefont{K.}~\bibnamefont{Leskovar}},
  \bibinfo{author}{\bibfnamefont{K.}~\bibnamefont{Staszek}},
  \bibinfo{author}{\bibfnamefont{M.}~\bibnamefont{F{\"o}lser}},
  \bibinfo{author}{\bibfnamefont{J.}~\bibnamefont{Weghuber}},
  \bibinfo{author}{\bibfnamefont{H.}~\bibnamefont{Stockinger}},
  \bibnamefont{and} \bibinfo{author}{\bibfnamefont{G.~J.}
  \bibnamefont{Sch{\"u}tz}}, \bibinfo{journal}{Nature methods}
  \textbf{\bibinfo{volume}{13}}, \bibinfo{pages}{661} (\bibinfo{year}{2016}).

\bibitem[{\citenamefont{Luke{\v{s}} et~al.}(2017)\citenamefont{Luke{\v{s}},
  Glatzov{\'a}, Kv{\'\i}{\v{c}}alov{\'a}, Levet, Benda, Letschert, Sauer,
  Brdi{\v{c}}ka, Lasser, and Cebecauer}}]{lukevs2017NatComm}
\bibinfo{author}{\bibfnamefont{T.}~\bibnamefont{Luke{\v{s}}}},
  \bibinfo{author}{\bibfnamefont{D.}~\bibnamefont{Glatzov{\'a}}},
  \bibinfo{author}{\bibfnamefont{Z.}~\bibnamefont{Kv{\'\i}{\v{c}}alov{\'a}}},
  \bibinfo{author}{\bibfnamefont{F.}~\bibnamefont{Levet}},
  \bibinfo{author}{\bibfnamefont{A.}~\bibnamefont{Benda}},
  \bibinfo{author}{\bibfnamefont{S.}~\bibnamefont{Letschert}},
  \bibinfo{author}{\bibfnamefont{M.}~\bibnamefont{Sauer}},
  \bibinfo{author}{\bibfnamefont{T.}~\bibnamefont{Brdi{\v{c}}ka}},
  \bibinfo{author}{\bibfnamefont{T.}~\bibnamefont{Lasser}}, \bibnamefont{and}
  \bibinfo{author}{\bibfnamefont{M.}~\bibnamefont{Cebecauer}},
  \bibinfo{journal}{Nat. Commun.} \textbf{\bibinfo{volume}{8}},
  \bibinfo{pages}{1} (\bibinfo{year}{2017}).

\bibitem[{\citenamefont{Sieber et~al.}(2006)\citenamefont{Sieber, Willig,
  Heintzmann, Hell, and Lang}}]{sieber2006BJ}
\bibinfo{author}{\bibfnamefont{J.~J.} \bibnamefont{Sieber}},
  \bibinfo{author}{\bibfnamefont{K.~I.} \bibnamefont{Willig}},
  \bibinfo{author}{\bibfnamefont{R.}~\bibnamefont{Heintzmann}},
  \bibinfo{author}{\bibfnamefont{S.~W.} \bibnamefont{Hell}}, \bibnamefont{and}
  \bibinfo{author}{\bibfnamefont{T.}~\bibnamefont{Lang}},
  \bibinfo{journal}{Biophys. J.} \textbf{\bibinfo{volume}{90}},
  \bibinfo{pages}{2843} (\bibinfo{year}{2006}).

\bibitem[{\citenamefont{Ben-Tal and Honig}(1996)}]{ben1996BJ}
\bibinfo{author}{\bibfnamefont{N.}~\bibnamefont{Ben-Tal}} \bibnamefont{and}
  \bibinfo{author}{\bibfnamefont{B.}~\bibnamefont{Honig}},
  \bibinfo{journal}{Biophys. J.} \textbf{\bibinfo{volume}{71}},
  \bibinfo{pages}{3046} (\bibinfo{year}{1996}).

\bibitem[{\citenamefont{Lorent and Levental}(2015)}]{lorent2015structural}
\bibinfo{author}{\bibfnamefont{J.~H.} \bibnamefont{Lorent}} \bibnamefont{and}
  \bibinfo{author}{\bibfnamefont{I.}~\bibnamefont{Levental}},
  \bibinfo{journal}{Chemistry and physics of lipids}
  \textbf{\bibinfo{volume}{192}}, \bibinfo{pages}{23} (\bibinfo{year}{2015}).

\bibitem[{\citenamefont{Anbazhagan and
  Schneider}(2010)}]{anbazhagan2010membrane}
\bibinfo{author}{\bibfnamefont{V.}~\bibnamefont{Anbazhagan}} \bibnamefont{and}
  \bibinfo{author}{\bibfnamefont{D.}~\bibnamefont{Schneider}},
  \bibinfo{journal}{Biochimica Et Biophysica Acta (BBA)-Biomembranes}
  \textbf{\bibinfo{volume}{1798}}, \bibinfo{pages}{1899}
  (\bibinfo{year}{2010}).

\bibitem[{\citenamefont{Schmidt et~al.}(2008)\citenamefont{Schmidt, Guigas, and
  Weiss}}]{schmidt2008PRL}
\bibinfo{author}{\bibfnamefont{U.}~\bibnamefont{Schmidt}},
  \bibinfo{author}{\bibfnamefont{G.}~\bibnamefont{Guigas}}, \bibnamefont{and}
  \bibinfo{author}{\bibfnamefont{M.}~\bibnamefont{Weiss}},
  \bibinfo{journal}{Phys. Rev. Lett.} \textbf{\bibinfo{volume}{101}},
  \bibinfo{pages}{128104} (\bibinfo{year}{2008}).

\bibitem[{\citenamefont{Milovanovic et~al.}(2015)\citenamefont{Milovanovic,
  Honigmann, Koike, G{\"o}ttfert, P{\"a}hler, Junius, M{\"u}llar, Diederichsen,
  Janshoff, Grubm{\"u}ller et~al.}}]{milovanovic2015NatComm}
\bibinfo{author}{\bibfnamefont{D.}~\bibnamefont{Milovanovic}},
  \bibinfo{author}{\bibfnamefont{A.}~\bibnamefont{Honigmann}},
  \bibinfo{author}{\bibfnamefont{S.}~\bibnamefont{Koike}},
  \bibinfo{author}{\bibfnamefont{F.}~\bibnamefont{G{\"o}ttfert}},
  \bibinfo{author}{\bibfnamefont{G.}~\bibnamefont{P{\"a}hler}},
  \bibinfo{author}{\bibfnamefont{M.}~\bibnamefont{Junius}},
  \bibinfo{author}{\bibfnamefont{S.}~\bibnamefont{M{\"u}llar}},
  \bibinfo{author}{\bibfnamefont{U.}~\bibnamefont{Diederichsen}},
  \bibinfo{author}{\bibfnamefont{A.}~\bibnamefont{Janshoff}},
  \bibinfo{author}{\bibfnamefont{H.}~\bibnamefont{Grubm{\"u}ller}},
  \bibnamefont{et~al.}, \bibinfo{journal}{Nat. Commun.}
  \textbf{\bibinfo{volume}{6}}, \bibinfo{pages}{1} (\bibinfo{year}{2015}).

\bibitem[{\citenamefont{West et~al.}(2009)\citenamefont{West, Brown, and
  Schmid}}]{west2009BJ}
\bibinfo{author}{\bibfnamefont{B.}~\bibnamefont{West}},
  \bibinfo{author}{\bibfnamefont{F.~L.} \bibnamefont{Brown}}, \bibnamefont{and}
  \bibinfo{author}{\bibfnamefont{F.}~\bibnamefont{Schmid}},
  \bibinfo{journal}{Biophys. J.} \textbf{\bibinfo{volume}{96}},
  \bibinfo{pages}{101} (\bibinfo{year}{2009}).

\bibitem[{\citenamefont{McMahon and Gallop}(2005)}]{mcmahon2005nature}
\bibinfo{author}{\bibfnamefont{H.~T.} \bibnamefont{McMahon}} \bibnamefont{and}
  \bibinfo{author}{\bibfnamefont{J.~L.} \bibnamefont{Gallop}},
  \bibinfo{journal}{Nature} \textbf{\bibinfo{volume}{438}},
  \bibinfo{pages}{590} (\bibinfo{year}{2005}).

\bibitem[{\citenamefont{Reynwar et~al.}(2007)\citenamefont{Reynwar, Illya,
  Harmandaris, M{\"u}ller, Kremer, and Deserno}}]{reynwar2007nature}
\bibinfo{author}{\bibfnamefont{B.~J.} \bibnamefont{Reynwar}},
  \bibinfo{author}{\bibfnamefont{G.}~\bibnamefont{Illya}},
  \bibinfo{author}{\bibfnamefont{V.~A.} \bibnamefont{Harmandaris}},
  \bibinfo{author}{\bibfnamefont{M.~M.} \bibnamefont{M{\"u}ller}},
  \bibinfo{author}{\bibfnamefont{K.}~\bibnamefont{Kremer}}, \bibnamefont{and}
  \bibinfo{author}{\bibfnamefont{M.}~\bibnamefont{Deserno}},
  \bibinfo{journal}{Nature} \textbf{\bibinfo{volume}{447}},
  \bibinfo{pages}{461} (\bibinfo{year}{2007}).

\bibitem[{\citenamefont{Goulian et~al.}(1993)\citenamefont{Goulian, Bruinsma,
  and Pincus}}]{goulian1993EPL}
\bibinfo{author}{\bibfnamefont{M.}~\bibnamefont{Goulian}},
  \bibinfo{author}{\bibfnamefont{R.}~\bibnamefont{Bruinsma}}, \bibnamefont{and}
  \bibinfo{author}{\bibfnamefont{P.}~\bibnamefont{Pincus}},
  \bibinfo{journal}{EPL (Europhysics Letters)} \textbf{\bibinfo{volume}{22}},
  \bibinfo{pages}{145} (\bibinfo{year}{1993}).

\bibitem[{\citenamefont{Park and Lubensky}(1996)}]{park1996JPI}
\bibinfo{author}{\bibfnamefont{J.-M.} \bibnamefont{Park}} \bibnamefont{and}
  \bibinfo{author}{\bibfnamefont{T.}~\bibnamefont{Lubensky}},
  \bibinfo{journal}{J. Phys. I} \textbf{\bibinfo{volume}{6}},
  \bibinfo{pages}{1217} (\bibinfo{year}{1996}).

\bibitem[{\citenamefont{Machta et~al.}(2012)\citenamefont{Machta, Veatch, and
  Sethna}}]{Machta12PRL}
\bibinfo{author}{\bibfnamefont{B.~B.} \bibnamefont{Machta}},
  \bibinfo{author}{\bibfnamefont{S.~L.} \bibnamefont{Veatch}},
  \bibnamefont{and} \bibinfo{author}{\bibfnamefont{J.~P.}
  \bibnamefont{Sethna}}, \bibinfo{journal}{Phys. Rev. Lett.}
  \textbf{\bibinfo{volume}{109}}, \bibinfo{pages}{138101}
  (\bibinfo{year}{2012}).

\end{thebibliography}

\clearpage 

\setcounter{table}{0}
\renewcommand{\thetable}{S\arabic{table}}%
\setcounter{figure}{0}
\renewcommand{\thefigure}{S\arabic{figure}}%
\setcounter{equation}{0}
\renewcommand{\theequation}{S\arabic{equation}}%

\begin{figure}
	\includegraphics[width=0.9\linewidth]{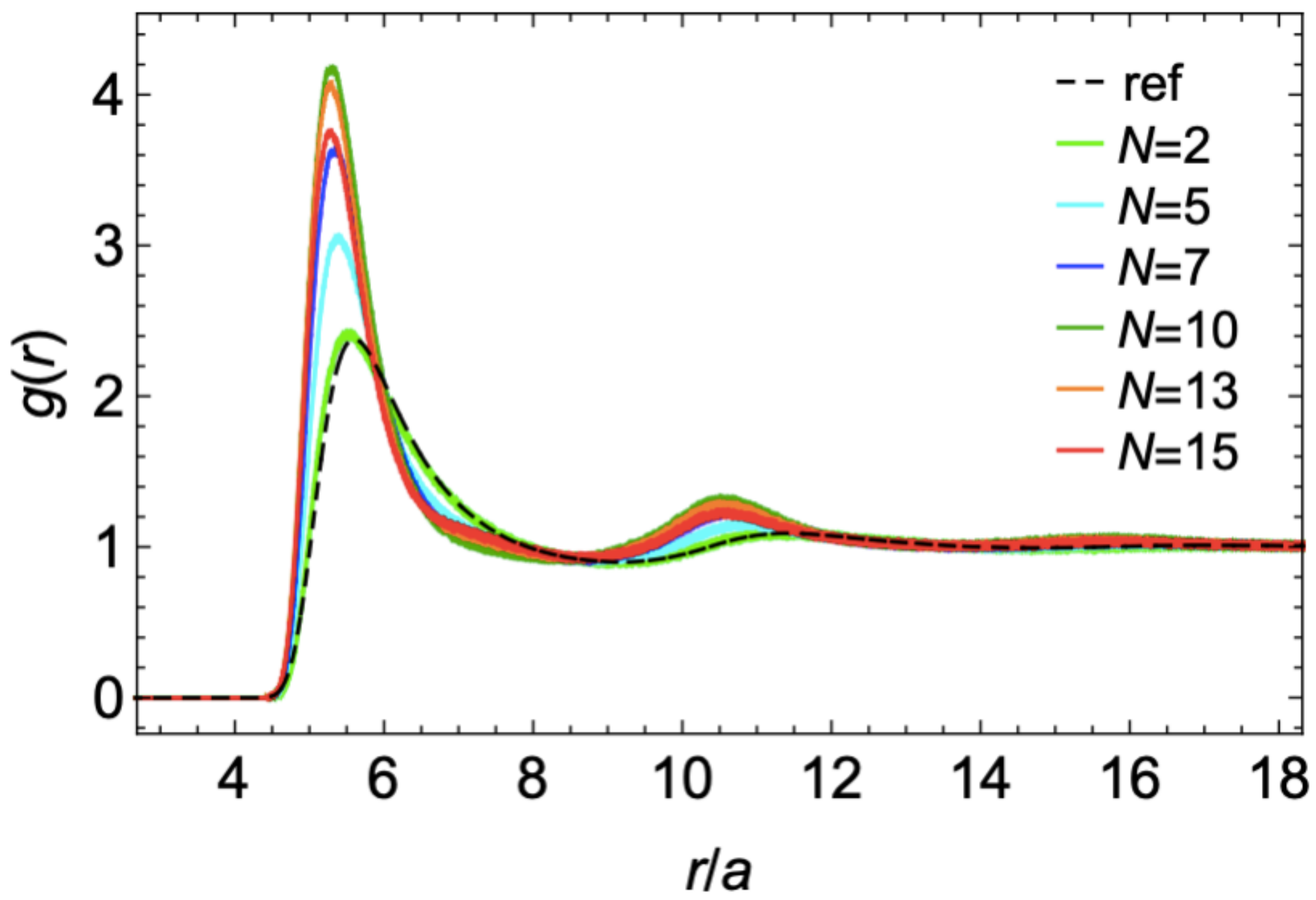}	\caption{\label{fig:g_r} The radial distribution function, $g(r)$, between the proteins for different brush sizes ($N$) with $\sigma a^2 = 0.09$, $\sigma_\text{P} a^2 = 0.01$ and $\beta \epsilon_\text{PP}=1$. 
	}
\end{figure}

\begin{figure}
	\includegraphics[width=0.8\linewidth]{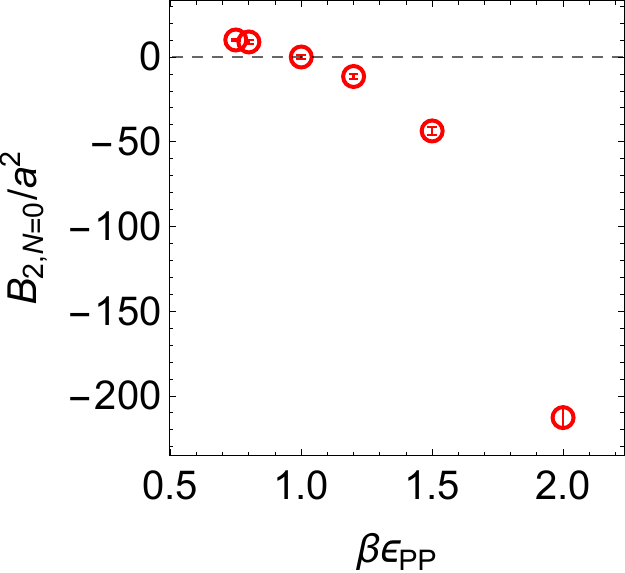}	\caption{\label{fig:B2_N0} The second virial coefficient, $B_{2, N=0}$, for the protein-only systems as a function of the interaction strength $\beta \epsilon_\text{PP}$ between the proteins with $\sigma_\text{P} a^2 = 0.01$, where $B_2^\text{ref}$ is depicted at $\beta \epsilon_\text{PP}=1$.
		}
\end{figure}

\begin{figure}	
       \includegraphics[width=0.8\linewidth]{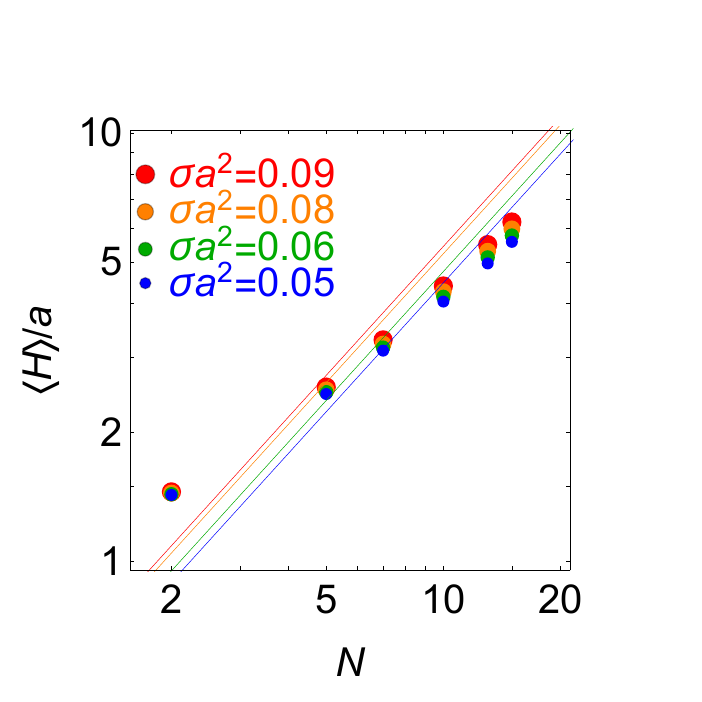}	
	\caption{The mean brush height, $\langle H \rangle$, as a function of the brush size ($N$) for different grafting densities ($\sigma$), shown in log-log scales. The solid lines depict the scaling relation $H=N \sigma^{1/3} b^{5/3}$ (Eq.~\eqref{eqn:brush_height}).}
\label{fig:meanH}
\end{figure}

\end{document}